\documentclass[aps,prd,amsmath,twocolumn,amssymbaps,showpacs]{revtex4-1}
\usepackage{graphicx}  
\usepackage{dcolumn}   
\usepackage{bm}        
\usepackage{amssymb}   
\usepackage{amsmath}   
\usepackage{bbm}
\usepackage{draftcopy}
\usepackage{relsize} 
\usepackage{xfrac}    
\usepackage{slashed}
\usepackage{color}
\usepackage{footnote}
\definecolor{dgreen}{cmyk}{1.,0.,1.,0.2}        
\definecolor{orange}{cmyk}{0.,0.353,1.,0.}    

\usepackage[bookmarks]{hyperref}

\newcommand{\di}{{\rm d}}

\newcommand{\tr}{{\rm tr}}

\newcommand{\be}{\begin{equation}}
\newcommand{\ee}{\end{equation}}                                                                               
\newcommand{\bea}{\begin{eqnarray}}
\newcommand{\eea}{\end{eqnarray}}

\begin{document}
\title{First-order QCD transition in a primordial magnetic field}

\author{Gaoqing Cao}
\affiliation{School of Physics and Astronomy, Sun Yat-sen University, Zhuhai 519088, China}

\date{\today}

\begin{abstract}
Recalling the expectation of an extremely strong primordial magnetic field $H$, we recheck transitions among the phases of chiral symmetry restoration ($\chi SR$), chiral symmetry breaking ($\chi SB$), and pion superfluidity ($\pi SF$)in the QCD epoch of the early universe. For homogeneous phases in a finite $H$, a sensible scheme is adopted to determine the phase boundaries of $\pi SF$, which is also superconductivity phase itself. In the first part, the QCD phase diagrams are studied in detail within the chiral effective Polyakov--Nambu--Jona-Lasinio model and the transitions involving $\pi SF$ are found to be of first order at relatively small $H$.  As expected from the Meissner effect, the regime of $\pi SF$ shrinks with increasing $H$ and completely vanishes  beyond a threshold value. In the second part, the bubble dynamics is illuminated for the stronger first-order transition, $\chi SR\rightarrow\pi SF$, in the more convenient Polyakov-quark-meson model. The coupled equations of motion of pion condensate and magnetic field are solved consistently to give the bubble structure. Then, based on bubble collisions, we explore gravitational wave (GW) emission by developing a simple toy model in advance; and the characteristic frequency of the relic GW is estimated to be of the order $0.1$--$1\,{\rm K}$ or $10^9$--$10^{10}\,{\rm Hz}$ in our galaxy.
\end{abstract}


\maketitle

\section{Introduction}
The exploration of possible phases in quantum chromodynamics (QCD) systems is a renewing topic in both low and high energy nuclear physics. In low energy nuclear physics, quarks are confined and the color singlet hadrons, mainly nucleons and pions, are expected to be effective degrees of freedom in nuclear matter. Due to strong attractive interactions, a dilute nucleonic system was found to prefer self-clustering to the saturation density with almost constant energy per nucleon~\cite{Weizsacker:1935bkz,Hofstadter:1956qs}. Such a property implies a first-order gas-liquid transition for isospin symmetric nuclear matter at low temperature~\cite{Fetter2003a}. Moreover, celestial neutron stars were expected to be realistic correspondences of isospin asymmetric nuclear matter~\cite{Li:2008gp}, and physicists had proposed many relevant phases for neutron stars, such as the one with the presence of Cooper pairing of neutrons~\cite{Elgaroy:1996mg}, pasta structure~\cite{Ravenhall:1983uh,Hashimoto:1984pap}, hyperon degrees of freedom~\cite{Glendenning:1992vb}, pion condensation~\cite{Akmal:1997ft}, or Kaon condensation~\cite{Lee:1996ef}, see also the review~\cite{Heiselberg:1999mq}. 

The high energy nuclear physics is characterized by considering quark degrees of freedom in the many-body QCD system~\cite{Lee:1974ma}. Relativistic heavy ion collisions~\cite{Lee:1974kn} were proposed to look for quark-gluon plasma, the phase where quarks and gluons are released from the interiors of nucleons. Such a phase was justified and its properties were also well studied in heavy ion colliders (HICs)~\cite{Yagi:2005yb}. In early stage, HICs usually explored high temperature and low chemical potential region where no sign of ordered phase transition was ever found~\cite{Aoki:2006we,Bhattacharya:2014ara,Floris:2014pta,Adamczyk:2017iwn}. Recently, several experiments focus on the low temperature and high chemical potential region in order to look for the critical end point (CEP) of chiral transition~\cite{Luo:2017faz}. Similarly, deconfined quarks were also proposed to exist in celestial bodies, mainly the inner cores of neutron stars~\cite{Baym:2017whm}and quark stars~\cite{Witten:1984rs}, with color superconductors of several kinds~\cite{Alford:2007xm} and quarkyonic matter~\cite{Fukushima:2015bda,McLerran:2018hbz,Cao:2020byn,Cao:2022inx} possible phases. 

Actually, the early universe is full of phase transitions~\cite{Baumann:2022mni} with electroweak (EW) and QCD transitions among the earliest ones. The EW transition was sometimes taken to account for baryogenesis~\cite{Trodden:1998ym} and strong primordial magnetic field was also assumed to be seeded there~\cite{Vachaspati:1991nm,Son:1998my,Grasso:2000wj}. Though nonvanishing, the baryon density is very small in the early universe~\cite{Planck:2015fie}, so the QCD chiral transition was expect be a trivial crossover, the same as that in HICs. However, the QCD epoch became more and more interesting when charged pion superfluidity was found to be favored for relatively large light lepton densities~\cite{Vovchenko:2020crk,Middeldorf-Wygas:2020glx,Cao:2021gfk}. The work is an extension of our previous work~\cite{Cao:2021gfk} by taking the primordial magnetic field into account. Two changes are expected: the second-order phase boundary of pion superfluidity might become of first order due to the Meissner effect induced by magnetic field~\cite{Fetter2003b}, and consequently the first-order transition would induce generation of gravitational wave~\cite{Witten:1984rs,Hogan:1986qda} even without inflation effect. Though GW emission has been widely studied in the first-order EW transition~\cite{Kosowsky:1992rz,Child:2012qg,Lewicki:2020azd,Wei:2022poh}, this might be the first time that a reliable first-order transition is applied to generate GW directly in the QCD epoch.

It is subtle to explore the possibility of superconductor phase when external magnetic field is present. For a type-II superconductor, vortical structure can be obtained by consistently solving the coupled equations of motion (EoMs) of charged condensate and magnetic field, and finite magnetic field could penetrate through the vortices~\cite{Pippard1953}. However, for a type-I superconductor, the magnetic field can only be present at the surface of the superconductor, known as the Meissner effect~\cite{London1935}. Then, how could we consistently explore the transition between phases with and without magnetic field in the bulk for a type-I superconductor? The problem had been well addressed in the textbook Ref.~\cite{Fetter2003b}: the external magnetic field ($H$), rather than the total magnetic field ($B$), is the same for both phases and thus serves the correct variable of free energy for the study of superconductivity. That means we have to transform the Helmholtz free energy with $B$ the variable to Gibbs free energy with $H$ the variable, and the Gibbs free energy of the superconductor can be shown to be the same as that with $H=0$.

The paper is mainly composed of two parts. In the first part, Sec.\ref{phase}, we explore the QCD phase diagrams of the early universe by accounting for primordial magnetic field within Polyakov--Nambu--Jona-Lasinio (PNJL) model.  Formalisms are developed for chiral phases and pion superfluidity in Sec.\ref{CP} and Sec.\ref{pion superfluidity}, separately, where we derive free energy, gap equations, and relevant  thermodynamic quantities. The most important sectors of strong interaction are studied in detail in Sec.\ref{PNJLB} and Sec.\ref{PNJL}, and the sectors of electroweak interaction are briefly summarized in Sec.\ref{EWB} and Sec.\ref{EW}. The numerical results of this part are shown in Sec.\ref{numerical} together with relevant discussions. In the second part, Sec.\ref{FOPT}, we firstly study bubble dynamics during the first-order QCD transition of the early universe by adopting the two-flavor Polyakov-quark-meson (PQM) model in Sec.\ref{BD}. And then in Sec.\ref{GWTS}, the results are applied to briefly explore the features of gravitational wave generated by bubble collisions within a toy model. Finally, an overall summary is given in Sec.\ref{summary}.
\section{Part I: phase diagrams with the three-flavor PNJL model}\label{phase}
In this section, we adopt the three-flavor PNJL model~\cite{Fukushima:2017csk,Klevansky:1992qe,Hatsuda:1994pi} for the QCD sector and explore in detail the phase diagrams with the presences of lepton flavor asymmetries and primordial magnetic field.
\subsection{Chiral phases in the magnetic field}\label{CP}
Usually, chiral symmetry restoration and breaking are related to the expectation value of one order parameter, i.e. the scalar field condensate. In the following, we specifically refer to the phases with only scalar field condensation as {\it chiral phases} to distinguish from the superconducting pion superfluidity phase where chiral symmetry is actually also broken. Due to the Meissner effect, the chiral phases and pion superfluidity should be treated separately in a background magnetic field.
\subsubsection{The strong interaction sector}\label{PNJLB}
In a primordial magnetic field, the Lagrangian of the PNJL model can be modified from the previous one~\cite{Cao:2021gfk} by adopting the covariant derivative $D_\mu=\partial_\mu+i\,Q_{\rm q}eA_\mu$ to
\begin{eqnarray}
{\cal L}_{\rm PNJL}\!&=&-{B^2\over 2}\!+\!\bar\psi\!\left[i\slashed{D}\!-\!i\gamma^4\!\!\left(\!ig{\cal A}^4\!+\!Q_{\rm q}\mu_{\rm Q}\!+\!{\mu_{\rm B}\over3}\!\right)\!-\!m_0\right]\!\psi\nonumber\\
&&\!\!\!+G\sum_{a=0}^8\!\left[(\bar\psi\lambda^a\psi)^2\!+\!(\bar\psi i\gamma_5\lambda^a\psi)^2\right]\!+\!{\cal L}_{\rm tH}\!-\!V(L).\label{LH}
\end{eqnarray}
Here, the field variables are defined as the following: $B$ is the magnetic field, $A_\mu$ is the corresponding vector potential, $\psi=(u,d,s)^T$ is the three-flavor quark field, and the Polyakov loop is $L={1\over N_{\rm c}}\tr\,e^{ig\int\di x_4{\cal A}^4}$ with ${\cal A}^4=A^{\rm 4c}T^{\rm c}/2$ the non-Abelian gauge field. For the quarks, the current mass and electric charge number matrices are respectively
 \bea
 m_0&\equiv&{\rm diag}(m_{\rm 0u},m_{\rm 0d},m_{\rm 0s}),\nonumber\\
 Q_{\rm q}&\equiv&{\rm diag}(q_{\rm u},q_{\rm d},q_{\rm s})={1\over3}{\rm diag}(2,-1,-1);
 \eea
and the interaction index $\lambda^0=\sqrt{2/3}~\mathbbm{1}_3$ and $\lambda^i~(i=1,\dots,8)$ are Gell-Mann matrices in flavor space. For later use, the 't Hooft term, ${\cal L}_{\rm tH}\equiv-K\sum_{t=\pm}{\rm Det}~\bar\psi\Gamma^t\psi$, can be represented as
\bea
\!\!\!\!\!\!\!\!{\cal L}_{\rm tH}\!=\!-{K\over2}\sum_{t=\pm}\epsilon_{ijk}\epsilon_{imn}(\bar{\psi}^i\Gamma^t{\psi}^i)(\bar{\psi}^j\Gamma^t{\psi}^m)(\bar{\psi}^k\Gamma^t{\psi}^n)
\eea
with the interaction vertices $\Gamma^\pm=\mathbbm{1}_4\pm\gamma_5$ for right- and left-handed channels, respectively. Here, one should note the Einstein summation convention for the flavor indices $i,j,k,m,n$ and the correspondences between $1,2,3$ and $u,d,s$. The pure gluon potential was usually obtained by fitting to the lattice QCD data, and we have in saddle point approximation:
 \bea
{V(L)\over T^4}&=&-{1\over2}\left(3.51-{2.47\over\tilde{T}}+{15.2\over\tilde{T}^{2}}\right)L^2-{1.75\over\tilde{T}^{3}}\nonumber\\
&&\times\ln\left[1-6\,L^2+8\,L^3-3L^4\right],
\eea
where $\tilde{T}\equiv T/T_0$ is the reduced temperature with $T_0=0.27\,{\rm GeV}$~\cite{Fukushima:2017csk}.

Now, in the chiral phases, we only consider nonzero chiral condensations $\sigma_{\rm i}\equiv\langle\bar{\psi}^i{\psi}^i\rangle$ with $i$ flavor index, then the 't Hooft term ${\cal L}_6$ can be reduced to an effective four fermion interaction forms in Hartree approximation~\cite{Klevansky:1992qe}:
\begin{widetext}
\bea
{\cal L}_6^4&=&-{K\over2}\sum_{s=\pm}\epsilon_{ijk}\epsilon_{imn}\langle\bar{\psi}^i\Gamma^s{\psi}^i\rangle(\bar{\psi}^j\Gamma^s{\psi}^m)(\bar{\psi}^k\Gamma^s{\psi}^n)\nonumber\\
&=&-{K\over6}\Big\{2\sum_{\rm f=u,d,s}\sigma_{\rm f}(\bar{\psi}\lambda^0\psi)^2-3\sigma_s\sum_{i=1}^3(\bar{\psi}\lambda^i\psi)^2
-3\sigma_{\rm d}\sum_{i=4}^5(\bar{\psi}\lambda^i\psi)^2-3\sigma_{\rm u}\sum_{i=6}^7(\bar{\psi}\lambda^i\psi)^2+(\sigma_s\!-\!2\sigma_{\rm u}\!-\!2\sigma_{\rm d})(\bar{\psi}\lambda^8\psi)^2\nonumber\\
&&+\sqrt{2}(2\sigma_s\!-\!\sigma_{\rm u}\!-\!\sigma_{\rm d})(\bar{\psi}\lambda^0\psi)(\bar{\psi}\lambda^8\psi)-\sqrt{6}(\sigma_{\rm u}\!-\!\sigma_{\rm d})(\bar{\psi}\lambda^3\psi)(\bar{\psi}\lambda^0\psi-\sqrt{2}\bar{\psi}\lambda^8\psi)\Big\}-(\lambda^a\rightarrow i\lambda^a\gamma^5)
\eea
with $\epsilon_{ijk}$ the Levi-Civita symbol. So the reduced three-flavor Lagrangian density with only four fermion effective interactions is
\begin{eqnarray}
\!\!\!\!\!\!{\cal L}_{\rm PNJL}^4\!\!=\!-{B^2\over 2}\!-\!V(L)\!+\!\bar\psi\!\left[i\slashed{D}\!-\!i\gamma^4\left(\!ig{\cal A}^4\!+\!Q_{\rm q}\mu_{\rm Q}\!+\!{\mu_{\rm B}\over3}\!\right)\!-\!m_0\right]\!\psi\!+\!\!\!\sum_{a,b=0}^8\!\!\left[G_{ab}^-(\bar\psi\lambda^a\psi)(\bar\psi\lambda^b\psi)\!+\!G_{ab}^+(\bar\psi i\gamma_5\lambda^a\psi)(\bar\psi i\gamma_5\lambda^b\psi)\right]\!,
\end{eqnarray}
where the nonvanishing elements of the symmetric coupling matrices $G^\pm$ are given by~\cite{Klevansky:1992qe}
\begin{eqnarray}
&&G_{00}^\mp=G\mp {K\over3}\sum_{\rm f=u,d,s}\sigma_{\rm f},~G_{11}^\mp=G_{22}^\mp=G_{33}^\mp=G\pm {K\over2}\sigma_s,~G_{44}^\mp=G_{55}^\mp=G\pm {K\over2}\sigma_{\rm d},~G_{66}^\mp=G_{77}^\mp=G\pm {K\over2}\sigma_{\rm u},\nonumber\\
&&G_{88}^\mp=G\mp {K\over6}(\sigma_s-2\sigma_{\rm u}-2\sigma_{\rm d}),~G_{08}^\mp=\mp {\sqrt{2}K\over12}(2\sigma_s\!-\!\sigma_{\rm u}\!-\!\sigma_{\rm d}),~G_{38}^\mp=-\sqrt{2}G_{03}^\mp=\mp {\sqrt{3}K\over6}(\sigma_{\rm u}\!-\!\sigma_{\rm d}).
\end{eqnarray}

By contracting a pair of field and conjugate field operators further in ${\cal L}_6^4$ in Hartree approximation, we find
\begin{eqnarray}
{\cal L}_6^2&=&-\sum_{s=\pm}^{i(\neq j\neq k)}K\langle\bar{\psi}^j\Gamma^s{\psi}^j\rangle\langle\bar{\psi}^k\Gamma^s{\psi}^k\rangle[\bar{\psi}^i\Gamma^s{\psi}^i]=-2K\sigma_j\sigma_k\bar{\psi}^i{\psi}^i\  (i\neq j\neq k, j<k),
\end{eqnarray}
which then, together with the contributions from the initial four-quark interactions, gives the effective quark masses as
\begin{eqnarray}\label{massi}
m_i&=&m_{0i}-4G\sigma_i+2K\sigma_j\sigma_k.
\end{eqnarray}
In order to evaluate quark masses numerically, we should be equipped with the gap equations directly following the definitions of chiral condensations:
\begin{eqnarray}
\sigma_i\equiv\langle\bar{\psi}^i{\psi}^i\rangle=-{i\over V_4}{\rm Tr}~{\cal S}_{i},
\end{eqnarray}
where the effective quark propagators in a constant magnetic field are given by~\cite{Cao:2021rwx}
\begin{eqnarray}
\hat{{\cal S}}_{\rm i}({k})
&=&i\int {\di s}\exp\Big\{-i (m_{\rm i}^{2}+{k}_4^2+k_3^2)s-i{\tan(q_{\rm i}eBs)\over q_{\rm i}eB}(k_1^2+k_2^2)\Big\}\left[m_{\rm i}-\gamma^4k_4\!-\!\gamma^3k_3\!-\!\gamma^2(k_2+{\tan(q_{\rm i}eBs)}k_1)\right.\nonumber\\
&&\left.-\gamma^1(k_1-{\tan(q_{\rm i}eBs)}k_2)\right]\Big[1
+{\gamma^1\gamma^2\tan(q_{\rm i}eBs)}\Big].\label{prop_m}
\end{eqnarray}

Then, by adopting vacuum regularization, the gap equations are~\cite{Cao:2021rwx}
\begin{eqnarray}
-\sigma_{\rm f}
&=&N_c{m_{\rm f}^3\over2\pi^2}\Big[\tilde{\Lambda}_{\rm f}\Big({1+\tilde{\Lambda}_{\rm f}^2}\Big)^{1\over2}-\ln\Big({\tilde\Lambda_{\rm f}}
+\Big({1+\tilde{\Lambda}_{\rm f}^2}\Big)^{1\over2}\Big)\Big]+N_c{m_{\rm f}\over4\pi^2}\int_0^\infty {ds\over s^2}e^{-m_{\rm f}^2s}\left({q_{\rm f}eBs
	\over\tanh(q_{\rm f}eBs)}-1\right)\nonumber\\
&&-6\sum_{\rm u=\pm}{|q_{\rm f}eB|\over 2\pi}\sum_{n=0}^\infty\alpha_{\rm n}\int_{-\infty}^\infty {\di  k_3\over2\pi}{m_{\rm f}
\over E_{\rm f}^{\rm n}}F_{\rm f}^{\rm u}(E_{\rm f}^{\rm n},L,T,\mu_{\rm Q},\mu_{\rm B}),\label{mgap}
\end{eqnarray}
where the reduced cutoff $\tilde{\Lambda}_{\rm f}={\Lambda/m_{\rm f}}$, Landau level factor $\alpha_{\rm n}=1-\delta_{\rm n0}/2$, particle energy $E_{\rm f}^{\rm n}(k_3,m_{\rm f})=(2n|q_{\rm f}eB|+k_3^2+m_{\rm f}^2)^{1/2}$, and the fermion distribution function $$F_{\rm f}^{\rm u}(E_{\rm f}^{\rm n},L,T,\mu_{\rm Q},\mu_{\rm B})\equiv{L\,e^{-{1\over T}\left(E_{\rm f}^{\rm n}-u\left(q_{\rm f}\mu_{\rm Q}+{\mu_{\rm B}\over 3}\right)\right)}+2L\,e^{-{2\over T}\left(E_{\rm f}^{\rm n}-u\left(q_{\rm f}\mu_{\rm Q}+{\mu_{\rm B}\over 3}\right)\right)}+e^{-{3\over T}\left(E_{\rm f}^{\rm n}-u\left(q_{\rm f}\mu_{\rm Q}+{\mu_{\rm B}\over 3}\right)\right)}\over 1+3L\,e^{-{1\over T}\left(E_{\rm f}^{\rm n}-u\left(q_{\rm f}\mu_{\rm Q}+{\mu_{\rm B}\over 3}\right)\right)}+3L\,e^{-{2\over T}\left(E_{\rm f}^{\rm n}-u\left(q_{\rm f}\mu_{\rm Q}+{\mu_{\rm B}\over 3}\right)\right)}+e^{-{3\over T}\left(E_{\rm f}^{\rm n}-u\left(q_{\rm f}\mu_{\rm Q}+{\mu_{\rm B}\over 3}\right)\right)}}.$$ In advance, the quark part of thermodynamic potential can be obtained consistently by combining the definitions of effective masses in Eq.\eqref{massi} and the integrations over ${m_{\rm f}}$ of Eq.\eqref{mgap} as~\cite{Cao:2023bmk}
	\begin{eqnarray}
	\Omega_{\rm q}(B)&=&2G\sum_{{\rm f}=u,d,s}\sigma_{\rm f}^2-4K\prod_{{\rm f}=u,d,s}\sigma_{\rm f}-N_c\sum_{{\rm f}=u,d,s}\left\{{m_{\rm f}^4\over8\pi^2}\Big[\tilde{\Lambda}_{\rm f}\Big(1+{2\tilde{\Lambda}_{\rm f}^2}\Big)\Big({1+{\tilde{\Lambda}_{\rm f}^2}}\Big)^{1\over2}-\ln\Big({\tilde{\Lambda}_{\rm f}}
+\Big({1+{\tilde{\Lambda}_{\rm f}^2}}\Big)^{1\over2}\Big)\Big]\right.\nonumber\\
&&-{1\over8\pi^2}\int_0^\infty {ds\over s^3}\left(e^{-m_{\rm f}^2s}-e^{-{m_{\rm f}^{\rm v}}^2s}\right)\left({q_{\rm f}eBs
	\over\tanh(q_{\rm f}eBs)}-1\right)-{1\over8\pi^2}\int_0^\infty {ds\over s^3}e^{-{m_{\rm f}^{\rm v}}^2s}\left({q_{\rm f}eBs
	\over\tanh(q_{\rm f}eBs)}-1-{1\over 3}(q_{\rm f}eBs)^2\right)\nonumber\\
&&\left.+2T\sum_{\rm u=\pm}{|q_{\rm f}eB|\over 2\pi}\sum_{n=0}^\infty\alpha_{\rm n}\int_{-\infty}^\infty {\di  k_3\over2\pi}K_{\rm f}^{\rm u}(E_{\rm f}^{\rm n},L,T,\mu_{\rm Q},\mu_{\rm B})\right\},\label{Omgq}
\end{eqnarray}
with
$$K_{\rm f}^{\rm u}(E_{\rm f}^{\rm n},L,T,\mu_{\rm Q},\mu_{\rm B})={1\over N_{\rm c}}\ln\left[1+3L\,e^{-{1\over T}\left(E_{\rm f}^{\rm n}-t\left(q_{\rm f}\mu_{\rm Q}+{\mu_{\rm B}\over 3}\right)\right)}+3L\,e^{-{2\over T}\left(E_{\rm f}^{\rm n}-t\left(q_{\rm f}\mu_{\rm Q}+{\mu_{\rm B}\over 3}\right)\right)}+e^{-{3\over T}\left(E_{\rm f}^{\rm n}-t\left(q_{\rm f}\mu_{\rm Q}+{\mu_{\rm B}\over 3}\right)\right)}\right].$$
Here, one notes that the terms depending on the quark vacuum mass $m_{\rm f}^{\rm v}$ are introduced for the correct renormalizations of electric charges and magnetic field in the vacuum. 

Hence, the Helmholtz free energy for the PNJL model is $\Omega_{\rm H}^{\rm M}={B^2\over 2}+V(L)+\Omega_{\rm q}(B)$ and the {\it external} magnetic field can be obtained through $H={\partial\Omega_{\rm H}^{\rm M}\over\partial B}$. In classical words,  the magnetic intensity $H$ equals magnetic induction intensity $B$ minus magnetization intensity ${\cal M}=-{\partial\Omega_{\rm q}\over\partial B}$. Usually, we control the external magnetic field $H$ for the exploration of phase transitions~\cite{Fetter2003b}, that is, $H$ must be a variable of the free energy. So, the right state function is the Gibbs free energy which can be obtained by taking Legendre transformation of $\Omega_{\rm H}^{\rm M}$ as~\cite{Fetter2003b} 
\bea
\Omega_{\rm \chi}^{\rm M}=\Omega_{\rm H}^{\rm M}-BH= -{H^2\over 2}+{{\cal M}^2\over 2}+V(L)+\Omega_{\rm q}(H+{\cal M}).\label{Gbs}
\eea
As the magnetizations from quarks and leptons are relatively small for the considered magnetic field in the chiral phases, we could simple take the Gibbs free energy to be $\Omega_{\rm \chi}^{\rm M}=-{H^2\over 2}+V(L)+\Omega_{\rm q}(H)$ to the order $o({\cal M}^2)$. The general formula Eq.\eqref{Gbs} even consistently applies to the superconducting pion superfluidity, where $B=H+{\cal M}=0$ in the bulk due to the Meissner effect; and we find $\Omega_{\pi SF}^{\rm M}=V(L)+\Omega_{\rm q}(0)$, just the same as the case without external magnetic field~\cite{Fetter2003b}. 

Then, the gap equation for $L$ can be given through $\partial_{\rm L}\Omega_{\rm \chi}^{\rm M}=0$ as
\bea
&&T^3\!\!\left[-\left(3.51\!-{2.47\over\tilde{T}}\!+\!{15.2\over\tilde{T}^{2}}\right)L\!+\!{1.75\over\tilde{T}^{3}}{12L(1-L)^2\over1\!-\!6L^2\!+\!8L^3\!-\!3L^4}\right]={6}\sum_{\rm f=u,d,s}\sum_{\rm u=\pm}{|q_{\rm f}eH|\over 2\pi}\sum_{n=0}^\infty\alpha_{\rm n}\int_{-\infty}^\infty {\di  k_3\over2\pi}\nonumber\\
&&{\,e^{-{1\over T}\left(E_{\rm f}^{\rm n}-u\left(q_{\rm f}\mu_{\rm Q}+{\mu_{\rm B}\over 3}\right)\right)}+\,e^{-{2\over T}\left(E_{\rm f}^{\rm n}-u\left(q_{\rm f}\mu_{\rm Q}+{\mu_{\rm B}\over 3}\right)\right)}\over 1+3L\,e^{-{1\over T}\left(E_{\rm f}^{\rm n}-u\left(q_{\rm f}\mu_{\rm Q}+{\mu_{\rm B}\over 3}\right)\right)}+3L\,e^{-{2\over T}\left(E_{\rm f}^{\rm n}-u\left(q_{\rm f}\mu_{\rm Q}+{\mu_{\rm B}\over 3}\right)\right)}+e^{-{3\over T}\left(E_{\rm f}^{\rm n}-u\left(q_{\rm f}\mu_{\rm Q}+{\mu_{\rm B}\over 3}\right)\right)}}.
\eea
And the entropy, electric charge number, and baryon number densities follow the thermodynamic relations as
\bea
\!\!\!\!\!\!s_{\rm \chi}^{\rm M}&=&2N_{\rm c}\sum_{\rm f=u,d,s}\sum_{\rm u=\pm}{|q_{\rm f}eH|\over 2\pi}\sum_{n=0}^\infty\alpha_{\rm n}\int_{-\infty}^\infty {\di  k_3\over2\pi}\left[K_{\rm f}^{\rm u}(E_{\rm f}^{\rm n},L,T,\mu_{\rm Q},\mu_{\rm B})+{1\over T}\left(E_{\rm f}^{\rm n}-u\left(q_{\rm f}\mu_{\rm Q}+{\mu_{\rm B}\over 3}\right)\right)F_{\rm f}^{\rm u}(E_{\rm f}^{\rm n},L,T,\mu_{\rm Q},\mu_{\rm B})\right]\nonumber\\
&&+T^3\left\{{1\over2}\left(4\times3.51-3\times{2.47\over\tilde{T}}+2\times{15.2\over\tilde{T}^{2}}\right)L^2+{1.75\over\tilde{T}^{3}}\ln\left[1-6L^2+8L^3-3L^4\right]\right\},\label{s_B}\\
\!\!\!\!\!\!n_{\rm Q}^{\rm q,M}&=&2N_{\rm c}\sum_{\rm f=u,d,s}\sum_{\rm u=\pm}{|q_{\rm f}eH|\over 2\pi}\sum_{n=0}^\infty\alpha_{\rm n}\int_{-\infty}^\infty {\di  k_3\over2\pi}u\,q_{\rm f}F_{\rm f}^{\rm u}(E_{\rm f}^{\rm n},L,T,\mu_{\rm Q},\mu_{\rm B}),\label{nQ_B}\\
\!\!\!\!\!\!n_{\rm B}^{\rm M}&=&2\sum_{\rm f=u,d,s}\sum_{\rm u=\pm}{|q_{\rm f}eH|\over 2\pi}\sum_{n=0}^\infty\alpha_{\rm n}\int_{-\infty}^\infty {\di  k_3\over2\pi}uF_{\rm f}^{\rm u}(E_{\rm f}^{\rm n},L,T,\mu_{\rm Q},\mu_{\rm B}).\label{nB_B}
\eea
\end{widetext}

\subsubsection{The electroweak interaction sector}\label{EWB}
In free gas approximation, the thermodynamic potentials for the quantum electroweak dynamics (QEWD) sector can be easily given by~\cite{Kapusta2006,Schwinger:1951nm}
\begin{widetext}
\bea
\Omega_\gamma&=&2T\int{\di^3k\over(2\pi)^3}\log\left(1-e^{- k/T}\right),\\
\Omega_{\rm l}^{\rm M}&=&\sum^{\rm i=e,\mu,\tau}\left\{-T\sum_{u=\pm}\int{\di^3k\over(2\pi)^3}\log\left[1+e^{- (k-u\,\mu_{\rm i})/T}\right]+{1\over8\pi^2}\int_0^\infty{\di s\over s^3}e^{-m_{\rm i}^2s}\left[{eHs\over\tanh(eHs)}-1-{1\over3}(eHs)^2\right]\right.\nonumber\\
&&\left.-2T\sum_{u=\pm}{|eH|\over 2\pi}\sum_{n=0}^\infty\alpha_{\rm n}\int_{-\infty}^\infty {\di  k_3\over2\pi}\log\left[1+e^{- \left(\epsilon_{\rm i}^{\rm n}(k_3,eH)-u\,(-\mu_{\rm Q}+\mu_{\rm i})\right)/T}\right]\right\},\label{OmegaBl}
\eea
where the degeneracy is one for neutrinos and anti-neutrinos due to their definite chiralities and $\epsilon_{\rm i}^{\rm n}(k_3,eH)=(k_3^2+2n|eH|+m_{\rm i}^2)^{1/2}$. Note that we have approximated $B$ by $H$ here. Then, the corresponding entropy, electric charge number and lepton flavor number densities can be derived directly as
\bea
s_\gamma&=&2\int{\di^3k\over(2\pi)^3}\left[-\log\left(1-e^{- k/T}\right)+{k/T\over e^{k/T}-1}\right],\\
s_{\rm l}^{\rm M}&=&\sum^{\rm i=e,\mu,\tau}_{u=\pm}\left\{\int{\di^3k\over(2\pi)^3}
\left\{\log\left[1+e^{- (k-u\,\mu_{\rm i})/T}\right]+{(k-u\,\mu_{\rm i})/T\over1+e^{(k-u\,\mu_{\rm i})/T}}\right\}+2\sum_{u=\pm}{|eH|\over 2\pi}\sum_{n=0}^\infty\alpha_{\rm n}\int_{-\infty}^\infty {\di  k_3\over2\pi}\right.\nonumber\\
&&\left.\left\{\log\left[1+e^{- \left(\epsilon_{\rm i}^{\rm n}(k_3,eH)-u\,(-\mu_{\rm Q}+\mu_{\rm i})\right)/T}\right]+{\left(\epsilon_{\rm i}^{\rm n}(k_3,eH)-u\,(-\mu_{\rm Q}+\mu_{\rm i})\right)/T\over1+e^{\left(\epsilon_{\rm i}^{\rm n}(k_3,eH)-u\,(-\mu_{\rm Q}+\mu_{\rm i})\right)/T}}\right\}\right\},\\
n_{\rm Q}^{\rm l,M}&=&2T\sum_{u=\pm}{|eH|\over 2\pi}\sum_{n=0}^\infty\alpha_{\rm n}\int_{-\infty}^\infty {\di  k_3\over2\pi}{-u\over1+e^{\left(\epsilon_{\rm i}^{\rm n}(k_3,eH)-u\,(-\mu_{\rm Q}+\mu_{\rm i})\right)/T}},\\
n_{\rm i}^{\rm M}&=&T\sum_{u=\pm}\int{\di^3k\over(2\pi)^3}{u\over1+e^{(k-u\,\mu_{\rm i})/T}}+2T\sum_{u=\pm}{|eH|\over 2\pi}\sum_{n=0}^\infty\alpha_{\rm n}\int_{-\infty}^\infty {\di  k_3\over2\pi}{u\over1+e^{\left(\epsilon_{\rm i}^{\rm n}(k_3,eH)-u\,(-\mu_{\rm Q}+\mu_{\rm i})\right)/T}}.
\eea
\end{widetext}

Now, collecting contributions from both the QEWD and QCD sectors, the total thermodynamic potential, entropy, electric charge number and lepton number densities are respectively
\bea
\Omega_{\rm M}&=&\Omega_\gamma+\Omega_{\rm l}^{\rm M}+\Omega_{\rm \chi}^{\rm M},\,
s_{\rm M}=s_\gamma\!+\!s_{\rm l}^{\rm M}\!+\!s_{\rm \chi}^{\rm M},\nonumber\\
n_{\rm Q}^{\rm M}&=&n_{\rm Q}^{\rm l,M}\!+\!n_{\rm Q}^{\rm q,M},\ \ \ \ \ \ \ \,
n_{\rm l}^{\rm M}=\!\sum_{\rm i=e,\mu,\tau} \!\! n_{\rm i}^{\rm M}\nonumber\\
\eea
in the QCD epoch. To better catch the expansion nature of the early universe, we define several reduced quantities:
\bea
b^{\rm M}=n_{\rm B}^{\rm M}/s_{\rm M},\ l^{\rm M}=n_{\rm l}^{\rm M}/s_{\rm M},\ l_{\rm i}^{\rm M}=n_{\rm i}^{\rm M}/s_{\rm M}
\eea
by following the conventions.
\subsection{The superconducting pion superfluidity}\label{pion superfluidity}
Since the pion superfluid is also an electric superconductor, the magnetic field will be screened from the bulk due to the Meissner effect. Then, the free energy of the bulk must be the same as the one with the same temperature and chemical potentials but without background magnetic field. So the formalism is the same as the one we presented in our previous work~\cite{Cao:2021gfk} where magnetic effect was not taken into account.
\subsubsection{The strong interaction sector}\label{PNJL}
 Without magnetic field, the Lagrangian is given by~\cite{Fukushima:2017csk,Klevansky:1992qe,Hatsuda:1994pi}:
\begin{eqnarray}
{\cal L}_{\rm PNJL}\!&=&\!-\!V(L)\!+\!\bar\psi\!\left[i\slashed{\partial}\!-\!i\gamma^4\!\!\left(\!ig{\cal A}^4\!+\!Q_{\rm q}\mu_{\rm Q}\!+\!{\mu_{\rm B}\over3}\!\right)\!-\!m_0\right]\!\psi\nonumber\\
&&+G\sum_{a=0}^8\left[(\bar\psi\lambda^a\psi)^2+(\bar\psi i\gamma_5\lambda^a\psi)^2\right]+{\cal L}_{\rm tH}.
\end{eqnarray}
For the pion superfluidity phase, we choose the following scalar and charged pseudoscalar condensates to be nonzero: $$\sigma_{\rm f}=\langle\bar\psi_{\rm f}\psi_{\rm f}\rangle,\ \Delta_\pi=\langle\bar{u}i\gamma^5d\rangle,\ \Delta_\pi^*=\langle\bar{d}i\gamma^5u\rangle.$$ For brevity, we set $\Delta_\pi=\Delta_\pi^*$ without loss of generality in the following. To facilitate the study, we'd like first to reduce ${\cal L}_{\rm tH}$ to an effective form with four-fermion interactions at most. By applying the Hartree approximation to contract a pair of quark and antiquark in each six-fermion interaction term~\cite{Klevansky:1992qe}, we immediately find
\begin{eqnarray}
{\cal L}_{\rm tH}^4
\!&=&\!-{K}\!\left\{\epsilon_{ijk}\epsilon_{imn}\sigma_i\!\left(\bar{\psi}^j{\psi}^m\bar{\psi}^k{\psi}^n
\!-\!\bar{\psi}^ji\gamma^5{\psi}^m\bar{\psi}^ki\gamma^5{\psi}^n\right)\!+\right.\nonumber\\
&&\left.\!\!\!\!2\Delta_\pi\!\!\left[\bar{s}{s}\!\left(\bar{u}i\gamma^5d\!+\!\bar{d}i\gamma^5u\!-\!\Delta_\pi\right)\!+\!\bar{s}i\gamma^5{s}\left(\bar{u}d\!+\!\bar{d}u\right)\right]\right\},
\end{eqnarray}
 where the second term in the brace is induced by $\pi^\pm$ condensations. Armed with the reduced Lagrangian density:
\begin{eqnarray}\label{LNJL4}
{\cal L}_{\rm PNJL}\!\!&=&\!\!-V(L,L)\!+\!\bar\psi\!\left[i\slashed{\partial}\!-\!i\gamma^4\!\!\left(\!ig{\cal A}^4\!+\!Q_{\rm q}\mu_{\rm Q}\!+\!{\mu_{\rm B}\over3}\!\right)\!-\!m_0\right]\!\psi\nonumber\\
&&+G\sum_{a=0}^8\left[(\bar\psi\lambda^a\psi)^2+(\bar\psi i\gamma_5\lambda^a\psi)^2\right]+{\cal L}_{\rm tH}^4,
\end{eqnarray}
the left calculations can just follow the two-flavor case in principle. 

By contracting quark and antiquark pairs once more in the interaction terms of Eq.\eqref{LNJL4}, we find the quark bilinear form as
\begin{eqnarray}\label{LNJL2}
{\cal L}_{\rm PNJL}^2\!\!=\!\bar\psi\left[i\slashed{\partial}\!-\! i\gamma^4\!\!\left(ig{\cal A}^4\!+\!Q_{\rm q}\mu_{\rm Q}\!+\!{\mu_{\rm B}\over3}\right)\!-\!m_{\rm i}\!-\!i\gamma^5\lambda^1\Pi \right]\psi,\nonumber\\
\end{eqnarray}
where the scalar and pseudoscalar masses are respectively
\begin{eqnarray}
m_i&=&m_{0i}-4G\sigma_i+2K(\sigma_j\sigma_k+\Delta_\pi^2\delta_{i3}),\nonumber\\
\Pi &=&(-4G+2K\sigma_3)\Delta_\pi\label{mpi}
\end{eqnarray}
with $i\neq j\neq k$. The $G$ and $K$ dependent terms in Eq.~\eqref{mpi} are from the $U_A(1)$ symmetric and anomalous interactions, respectively. According to Eq.\eqref{LNJL2}, $s$ quark decouples from $u,d$ quarks, so the gap equation for $\sigma_{\rm s}$ can be simply given by~\cite{Klevansky:1992qe}:
\begin{eqnarray}
\!\!\!\!\!\sigma_{\rm s}\!=\!\langle\bar{s}{s}\rangle\!=\!{\rm tr} \left[i\slashed{\partial}-i\gamma^4\!\!\left(ig{\cal A}^4\!+\!Q_{\rm q}\mu_{\rm Q}\!+\!{\mu_{\rm B}\over3}\right)\!-\!m_{\rm s} \right]^{\!-\!1}\!\!.\label{sigmas}
\end{eqnarray}
However, the $u$ and $d$ light quarks couple with each other through the non-diagonal pseudoscalar mass $\Pi$. Since $\mu_{\rm B}$ is usually small in early universe, we can simply set $$m_{\rm 0u}=m_{\rm 0d}\equiv m_{\rm 0l},\ \sigma_{\rm u}=\sigma_{\rm d}\equiv\sigma_{\rm l}$$ in order to further carry out analytic derivations. Then, by following a similar procedure as the previous section, the explicit thermodynamic potential can be worked out for the bilinear terms as
\begin{widetext}
\bea
\!\!\!\!\!\!\Omega_{\rm bl}&=&\!-\!2N_c\!\!\int^\Lambda\!\!\!\!{\di^3k\over(2\pi)^3}\!\!\left[\sum_{t=\pm}\!E_{\rm l}^{\rm t}(k)\!+\!\epsilon_{\rm s}(k)\right]\!\!-\!2T\!\!\int\!\!\!{\di^3k\over(2\pi)^3}\!\!\sum_{\rm u=\pm}\!\!\left[\sum_{t=\pm}\!Fl\!\left(\!L,u,E_{\rm l}^{\rm t}(k),{\mu_{\rm Q}\!+\!2\mu_{\rm B}\over6}\!\right)\!\!+\!Fl\!\left(\!L,u,\epsilon_{\rm s}(k),{-\mu_{\rm Q}\!+\!\mu_{\rm B}\over3}\!\right)\!\right],\\
&&\ \ \ \ \ \ \ \left\{Fl(L,u,x,y)=\log\left[1+3L\,e^{-{1\over T}\left(x-u\,y\right)}+3L\,e^{-{2\over T}\left(x-u\,y\right)}+e^{-{3\over T}\left(x-u\,y\right)}\right]\right\}\nonumber
\eea
\end{widetext}
with the particle energy functions defined by
\bea
\!\!\!\!\epsilon_{\rm i}(k)\!=\!\sqrt{k^2\!+m_{\rm i}^2},\
E_{\rm l}^{\rm t}(k)\!=\!\sqrt{\left[\epsilon_{\rm l}(k)\!+t{\mu_{\rm Q}\over2}\right]^2\!+\Pi^2}.
\eea
Eventually, the coupled gap equations follow directly from the definitions of condensates:
\bea
&&\sigma_{\rm s}\equiv\langle\bar{s}{s}\rangle={\partial\Omega_{\rm bl}\over\partial m_{\rm s}},\
2\sigma_{\rm l}\equiv\langle\bar{u}{u}\rangle+\langle\bar{d}{d}\rangle={\partial\Omega_{\rm bl}\over\partial m_{\rm l}},\nonumber\\
&&2\Delta_\pi\equiv\langle\bar{u}i\gamma^5{d}\rangle+\langle\bar{d}i\gamma^5{u}\rangle={\partial\Omega_{\rm bl}\over\partial \Pi}\label{conds}
\eea
and the minimal condition $\partial_{\rm L}[V(L,L)+\Omega_{\rm bl}]=0$ as~\cite{Cao:2021gfk}
\begin{widetext}
\bea
\sigma_{\rm s}
\!&=&\!-2N_c\int^\Lambda\!\!{\di^3k\over(2\pi)^3}{m_{\rm s}\over\epsilon_{\rm s}(k)}+2N_c\int\!\!{\di^3k\over(2\pi)^3}{m_{\rm s}\over\epsilon_{\rm s}(k)}\sum_{u=\pm}dV_1\left(\!L,u,\epsilon_{\rm s}(k),{-\mu_{\rm Q}\!+\!\mu_{\rm B}\over3}\!\right),\label{gaps}\\
2\sigma_{\rm l}\!&=&\!-2N_c\int^\Lambda\!\!{\di^3k\over(2\pi)^3}\sum_{t=\pm}{m_{\rm l}\over\epsilon_{\rm l}(k)}{\epsilon_{\rm l}(k)+t{\mu_{\rm Q}\over2}\over E_{\rm l}^{\rm t}(k)}+2N_c\int\!\!{\di^3k\over(2\pi)^3}\sum_{t,u=\pm}{m_{\rm l}\over\epsilon_{\rm l}(k)}{\epsilon_{\rm l}(k)+t{\mu_{\rm Q}\over2}\over E_{\rm l}^{\rm t}(k)}dV_1\left(\!L,u,E_{\rm l}^{\rm t}(k),{\mu_{\rm Q}\!+\!2\mu_{\rm B}\over6}\!\right),\label{gapl}\\
2\Delta_\pi\!&=&\!-2N_c\int^\Lambda\!\!{\di^3k\over(2\pi)^3}\sum_{t=\pm}{\Pi\over E_{\rm l}^{\rm t}(k)}+2N_c\int\!\!{\di^3k\over(2\pi)^3}\sum_{t,u=\pm}{\Pi\over E_{\rm l}^{\rm t}(k)}dV_1\left(\!L,u,E_{\rm l}^{\rm t}(k),{\mu_{\rm Q}\!+\!2\mu_{\rm B}\over6}\!\right),\label{gappi}
\eea
\bea
&&\!\!\!\!\!\!\!\!\!\!\!\!\!\!T^4\left[-\left(3.51\!-{2.47\over\tilde{T}}\!+\!{15.2\over\tilde{T}^{2}}\right)L\!+\!{1.75\over\tilde{T}^{3}}{12L(1-L)^2\over1\!-\!6L^2\!+\!8L^3\!-\!3L^4}\right]=6T\!\int\!\!{\di^3k\over(2\pi)^3}\!\sum_{u=\pm} \left[\sum_{t=\pm}dV_2\left(L,u,E^{\rm t}(k),{\mu_{\rm Q}\!+\!2\mu_{\rm B}\over6}\right)\right.\nonumber\\
&&\left.+dV_2\left(\!L,u,\epsilon_{\rm s}(k),{-\mu_{\rm Q}\!+\!\mu_{\rm B}\over3}\!\right)\right].
\eea
\end{widetext}
Note that $\Delta_\pi=0$ is a trivial solution of Eq.\eqref{gappi}, so $\Delta_\pi$ or $\Pi$ is still a true order parameter for $I_3$ isospin symmetry~\cite{Son:2000xc} in three-flavor case. The total self-consistent thermodynamic potential can be found to be
\bea
\Omega_{\rm \pi SF}&=&V(L,L)+\Omega_{\rm bl}+2G(\sigma_{\rm s}^2+2\sigma_{\rm l}^2+2\Delta_\pi^2)\nonumber\\
&&-4K(\sigma_{\rm l}^2+\Delta_\pi^2)\sigma_{\rm s}
\eea
by utilizing the definitions of condensates and their relations to scalar and pseudoscalar masses, refer to Eqs.\eqref{conds} and \eqref{mpi}. And the entropy, electric charge number and baryon number densities can be given according to the thermodynamic relations as~\cite{Cao:2021gfk}
\begin{widetext}
\bea
\!\!\!\!\!\!s_{\rm \pi SF}&=&2\!\!\int\!\!{\di^3k\over(2\pi)^3}\!\sum_{t,u=\pm}\left[Fl\left(\!L,u,E_{\rm l}^{\rm t}(k),{\mu_{\rm Q}\!+\!2\mu_{\rm B}\over6}\!\right)\!+\!{3\left(\!E_{\rm l}^{\rm t}(k)\!-\!u\,{\mu_{\rm Q}\!+\!2\mu_{\rm B}\over6}\!\right)\over T}dV_1\left(\!L,u,E_{\rm l}^{\rm t}(k),{\mu_{\rm Q}\!+\!2\mu_{\rm B}\over6}\!\right)\right]\nonumber\\
&&+2\!\!\int\!\!{\di^3k\over(2\pi)^3}\!\sum_{u=\pm}\left[Fl\left(\!L,u,\epsilon_{\rm s}(k),{-\mu_{\rm Q}\!+\!\mu_{\rm B}\over3}\!\right)\!+\!{3\left(\!E_{\rm l}^{\rm t}(k)\!-\!u\,{-\mu_{\rm Q}\!+\!\mu_{\rm B}\over3}\!\right)\over T}dV_1\left(\!L,u,\epsilon_{\rm s}(k),{-\mu_{\rm Q}\!+\!\mu_{\rm B}\over3}\!\right)\right]\nonumber\\
&&+T^3\left\{{1\over2}\left(4\times3.51-3\times{2.47\over\tilde{T}}+2\times{15.2\over\tilde{T}^{2}}\right)L^2+{1.75\over\tilde{T}^{3}}\ln\left[1-6L^2+8L^3-3L^4\right]\right\},\label{s3f}\\
\!\!\!\!\!\!n_{\rm Q}^{\rm \pi SF}&=&N_c\!\!\int^\Lambda\!\!{\di^3k\over(2\pi)^3}\!\sum_{t=\pm}t{\epsilon_{\rm l}(k)\!+\!t{\mu_{\rm Q}\over2}\over E_{\rm l}^{\rm t}(k)}-\!3\!\!\int\!\!{\di^3k\over(2\pi)^3}\sum_{t,u=\pm}t{\epsilon_{\rm l}(k)\!+\!t{\mu_{\rm Q}\over2}\over E_{\rm l}^{\rm t}(k)}dV_1\left(\!L,u,E_{\rm l}^{\rm t}(k),{\mu_{\rm Q}\!+\!2\mu_{\rm B}\over6}\!\right)\nonumber\\
&&+\int\!\!{\di^3k\over(2\pi)^3}\!\sum_{t,u=\pm}u\,dV_1\left(\!L,u,E_{\rm l}^{\rm t}(k),{\mu_{\rm Q}\!+\!2\mu_{\rm B}\over6}\!\right)-2\int{\di^3k\over(2\pi)^3}\!\sum_{t,u=\pm}u\,dV_1\left(\!L,u,\epsilon_{\rm s}(k),{-\mu_{\rm Q}\!+\!\mu_{\rm B}\over3}\!\right),\label{nQ3f}\\
\!\!\!\!\!\!n_{\rm B}^{\rm \pi SF}&=&2\int{\di^3k\over(2\pi)^3}\!\sum_{t,u=\pm}u\,dV_1\left(\!L,u,E_{\rm l}^{\rm t}(k),{\mu_{\rm Q}\!+\!2\mu_{\rm B}\over6}\!\right)+2\int{\di^3k\over(2\pi)^3}\!\sum_{t,u=\pm}u\,dV_1\left(\!L,u,\epsilon_{\rm s}(k),{-\mu_{\rm Q}\!+\!\mu_{\rm B}\over3}\!\right).\label{nB3f}
\eea
\end{widetext}

\subsubsection{The electroweak interaction sector}\label{EW}
In free gas approximation, the thermodynamic potentials for the QEWD sector can be easily given by~\cite{Kapusta2006}
\bea
\Omega_\gamma&=&2T\int{\di^3k\over(2\pi)^3}\log\left(1-e^{- k/T}\right),\\
\Omega_{\rm l}&=&-T\!\!\sum^{\rm i=e,\mu,\tau}_{u=\pm}\!\int\!\!{\di^3k\over(2\pi)^3}\!\left\{2\log\left[1+e^{- \left(\epsilon_{\rm i}(k)-u\,(-\mu_{\rm Q}+\mu_{\rm i})\right)/T}\right]\right.\nonumber\\
&&\left.+\log\left[1+e^{- (k-u\,\mu_{\rm i})/T}\right]\right\},\label{Omegal}
\eea
where the degeneracy is one for neutrinos and anti-neutrinos due to their definite chiralities. Then, the corresponding entropy, electric charge number and lepton flavor number densities can be derived directly as
\bea
s_\gamma&=&2\int{\di^3k\over(2\pi)^3}\left[-\log\left(1-e^{- k/T}\right)+{k/T\over e^{k/T}-1}\right],\\
s_{\rm l}&=&\sum^{\rm i=e,\mu,\tau}_{u=\pm}\int{\di^3k\over(2\pi)^3}\left\{2\log\left[1+e^{- \left(\epsilon_{\rm i}(k)-u\,(-\mu_{\rm Q}+\mu_{\rm i})\right)/T}\right]\right.\nonumber\\
&&+\log\left[1+e^{- (k-u\,\mu_{\rm i})/T}\right]\!+\!{2\left(\epsilon_{\rm i}(k)\!-\!u\,(\!-\mu_{\rm Q}\!+\!\mu_{\rm i})\right)/T\over1+e^{\left(\epsilon_{\rm i}(k)-u\,(-\mu_{\rm Q}+\mu_{\rm i})\right)/T}}\nonumber\\
&&\left.+{(k-u\,\mu_{\rm i})/T\over1+e^{(k-u\,\mu_{\rm i})/T}}\right\},
\eea
\bea
n_{\rm Q}^{\rm l}&=&2T\sum^{\rm i=e,\mu,\tau}_{u=\pm}\int{\di^3k\over(2\pi)^3}{-u\over1+e^{\left(\epsilon_{\rm i}(k)-u\,(-\mu_{\rm Q}+\mu_{\rm i})\right)/T}},\\
n_{\rm i}&=&-{\partial \Omega_{\rm l}\over \partial \mu_{\rm i}}=T\sum_{u=\pm}\int{\di^3k\over(2\pi)^3}\left[{2u\over1+e^{\left(\epsilon_{\rm i}(k)-u\,(-\mu_{\rm Q}+\mu_{\rm i})\right)/T}}\right.\nonumber\\
&&\qquad\qquad\ \ \ \left.+{u\over1+e^{(k-u\,\mu_{\rm i})/T}}\right],\ i=e,\mu,\tau.
\eea

Now, collecting contributions from both QEWD and QCD sectors, the total thermodynamic potential, entropy, electric charge number and lepton number densities are respectively
\bea
\Omega&=&\Omega_\gamma+\Omega_{\rm l}+\Omega_{\rm \pi SF},\,
s=s_\gamma\!+\!s_{\rm l}\!+\!s_{\rm \pi SF},\nonumber\\
n_{\rm Q}&=&n_{\rm Q}^{\rm l}\!+\!n_{\rm Q}^{\rm \pi SF},\ \ \ \ \ \ \ \,
n_{\rm l}=\!\sum_{\rm i=e,\mu,\tau} \!\! n_{\rm i}\nonumber\\
\eea
in the QCD epoch. To better catch the expansion nature of the early universe, we define several reduced quantities:
\bea
b=n_{\rm B}^{\rm \pi SF}/s,\ l=n_{\rm l}/s,\ l_{\rm i}=n_{\rm i}/s
\eea
by following the conventions. 

\subsection{Numerical results}\label{numerical}
To carry out numerical calculations, we get the electron and muon masses from the Particle Data Group as $m_{\rm e}=0.53\,{\rm MeV}$ and $m_\mu=113\,{\rm MeV}$ and suppress the contribution of heavy $\tau$ leptons for the electroweak interaction sector. The model parameters are fixed for the strong interaction sector as the following~\cite{Zhuang:1994dw,Rehberg:1995kh}
\bea
&& m_{\rm 0l}\!=\!5.5\,{\rm MeV},\, m_{\rm 0s}\!=\!140.7\,{\rm MeV},\, \Lambda\!=\!602.3\,{\rm MeV},\nonumber\\ 
&&G\Lambda^2\!=\!1.835,\, K\Lambda^5\!=\!12.36.
\eea

First of all, we have to determine which phase the QCD matter is in, the chiral phases or pion superfluidity, by comparing $\Omega^{\rm M}_{\chi}$ and $\Omega_{\pi SF}$. We choose the recent constraints $n^{\rm Q}=0, b^{\rm M}=8.6*10^{-11}$~\cite{Planck:2015fie} and $l^{\rm M}=-0.012$~\cite{Oldengott:2017tzj} up to the point when the latest first-order transition took place in the QCD epoch of the early universe. The non-constrained values of $l^{\rm M}_{\rm e}$ and $l^{\rm M}_{\rm \mu}$ can be randomly fixed at that point, but we only consider the case with $l^{\rm M}_{\rm e}=0$ for a given $l^{\rm M}_{\rm e}+l^{\rm M}_{\rm \mu}$. Note that our previous work had showed that the phase boundaries were not sensitive to the fraction of $l^{\rm M}_{\rm e}$~\cite{Cao:2021gfk}. When first-order phase transitions are involved, the total entropy does not change continuously at the transition point due to the latent heat released or absorbed, thus we should not require $b^{\rm M}=b$ or $l^{\rm M}_{\rm i}=l_{\rm i}, (i=e,\mu,\tau)$ in the bulk. For pion superfluidity, the magnetic field in the bulk is canceled out by the current produced at the surface and the magnetic flux is only present at the surface~\cite{Fetter2003b}. So in the sense of total baryon and lepton flavor number conservations, the deficits $n_{\rm B}^{\rm M}V^{\rm M}-n_{\rm b}V$ and $n_{\rm i}^{\rm M}V^{\rm M}-n_{\rm i}V$ should be found at the surface of the pion superfluidity. Nevertheless, the ratios $n_{\rm i}^{\rm M}/n_{\rm B}^{\rm M}$ must be the same in the $\chi SR$ and $\chi SB$ phases separated by pion superfluidity if exists. Following the ansatz of isentropic expansion for a given phase, we expect $n_{\rm B}^{\rm M}/s$ and $n_{\rm i}^{\rm M}/s$ to be the same at the entrance and exit of pion superfluidity. 

For the given values of $n^{\rm Q}, b^{\rm M}$, and $l^{\rm M}_{\rm i}$, the values of chemical potentials $\mu_{\rm Q}, \mu_{\rm B}$, and $\mu_{\rm i}$ can be uniquely determined in the chiral phases for a fixed temperature. And then the transition points could be pinned down by requiring $\Omega^{\rm M}_{\chi}=\Omega_{\pi SF}$ at the same temperature and chemical potentials. Now we can obtain two different sets of $b$ and $l_{\rm i}$ at the transition points, but how should they evolve in between in the pion superfluidity phase? For $H=0$, we expected them to be the same as the recent constraints since no extra particles are reserved at the surface. Any reasonable scheme must recover the results in the vanishing $H$ limit, hence we simply adopt linear interpolations between the two sets of $b$ and $l_{\rm i}$ for finite $H$. That means the baryons and lepton flavors are gradually deposited into or withdrawn from the surface with the temperature decreasing. 

Next, to study the effect of primordial magnetic field, the order of $H$ should be estimated in the QCD epoch of the early universe. In the Milky Way, the average magnitude of magnetic field was found to be $H=10^{-6}$--$10^{-5}\,{\rm Gs}$ according to the observations of galactic background radio radiation and polarization of star light~\cite{Verschuur1974}. In natural unit, the magnitude is $eH=10^{-26}$--$10^{-25}\,{\rm GeV}^2$ since $eH=5.9\times 10^{-21}\,{\rm GeV}^2$ for $H=1\,{\rm Gs}$. Tracing back to the QCD epoch, the magnetic field would be greatly enhanced due to a very large scaling factor $a=10^{12}$--$10^{12.5}$~\cite{Baumann:2022mni} and we have $eH=(10^{-26}$--$10^{-25})a^2=10^{-2}$--$1\,({\rm GeV}^2)$. As we will see, the favored region of pion superfluidity would greatly shrink with the magnetic field increasing. So in order to explore nontrivial physics, we focus on the lower region of the magnetic field domain estimated, that is, $eH\sim 10^{-2}\,{\rm GeV}^2$. Note that the magnetic field is not homogeneous all across the Milky Way but only locally, hence the volume where the homogeneous phase transition might happen should not be taken to be infinite. 

\begin{figure}[!htb]
	\begin{center}
	\includegraphics[width=8cm]{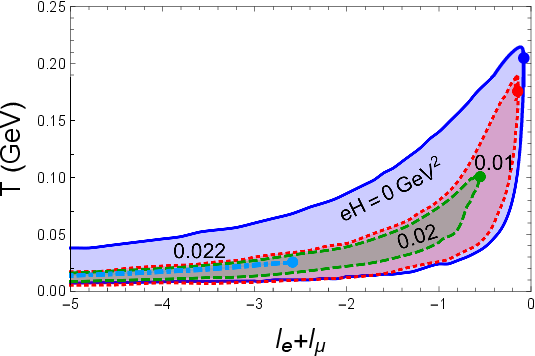}
	\includegraphics[width=8cm]{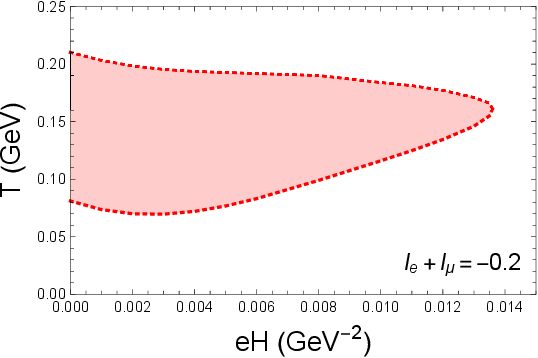}
		\caption{Upper panel: the $T-(l_{\rm e}^{\rm M}+l_\mu^{\rm M})$ phase diagrams for the magnetic fields $eH=0, 0.01, 0.02$, and $0.022\,{\rm GeV}^2$ (colors: blue, red, green, and cyan) with the bullets the critical end points. Lower panel: the $T-eH$ phase diagram for the lepton flavor-entropy ratio $l_{\rm e}^{\rm M}+l_\mu^{\rm M}=-0.2$. The shadows correspond to the pion superfluidity phase, and the blue dashed line and other colored lines denote second- and first-order transition boundaries, respectively.}\label{T-lB}
	\end{center}
\end{figure}
\begin{figure}[!htb]
	\begin{center}
		\includegraphics[width=8cm]{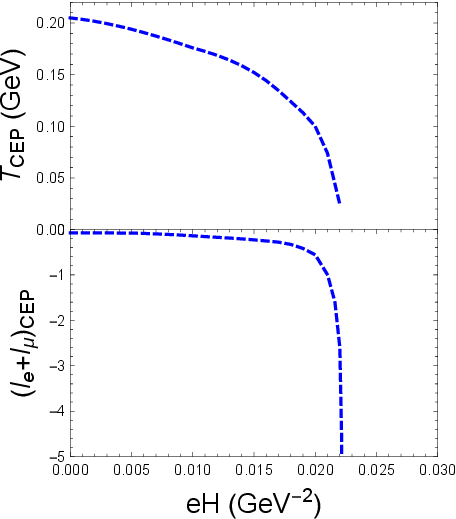}
		\caption{The temperature $T_{\rm CEP}$ (upper panel) and lepton flavor-entropy ratio $(l_{\rm e}^{\rm M}+l_\mu^{\rm M})_{\rm CEP}$ (lower panel) as functions of magnetic field $eH$ at the critical end points.}\label{CEP}
	\end{center}
\end{figure}

The phase diagrams with fixed background magnetic fields or lepton flavor densities are illustrated together in Fig.\ref{T-lB}. Since the densities are not continuous across the first-order transition point, we take the lepton flavor-entropy ratios $l_{\rm i}^{\rm M}$ in the chiral phases for reference. From the upper panel, one can tell that the regime of pion superfluidity shrinks quickly with increasing magnetic field but the tails are never find to end at a large $|l_{\rm e}^{\rm M}+l_\mu^{\rm M}|$. Nevertheless, when $eH$ exceeds $0.0222\,{\rm GeV}^2$, the width of the regime of the pion superfluidity is found to vanish and one is left with only chiral phases. This explains our focus on the magnitude of the magnetic field: $eH\sim 10^{-2}\,{\rm GeV}^2$. As shown in the lower panel, the phase boundaries become of first order and oscillate with finite $H$, which is known as the de Haas-van Alphen effect~\cite{Cao:2021rwx} when both magnetic field and chemical potentials are present. For small $H$, there are actually very special points where the transitions remain first order along $|l_{\rm e}^{\rm M}+l_\mu^{\rm M}|$ direction but are of second order along $T$ direction. These critical end points (CEPs) are actually where the upper and lower boundaries meet each other and thus with the smallest value of $|l_{\rm e}^{\rm M}+l_\mu^{\rm M}|$,  see the bullets in the upper panel of Fig.\ref{T-lB}. It is constructive to demonstrate the evolutions of the temperature $T$ and lepton flavor-entropy ratio $l_{\rm e}^{\rm M}+l_\mu^{\rm M}$ with increasing magnetic field $eH$ at the CEPs in Fig.\ref{CEP}. Consistent with the upper panel of Fig.\ref{T-lB},  $T_{\rm CEP}$ decreases gradually with $eH$ (upper panal) but $|l_{\rm e}^{\rm M}+l_\mu^{\rm M}|_{\rm CEP}$ increases abruptly around $eH=0.022\,{\rm GeV}^2$ (lower panel), which seem to be related to the transition orders just discussed. Moreover, from the lower panel of Fig.\ref{CEP}, we can easily understand the complete disfavor of pion superfluidity for larger $eH$. 

\begin{figure}[!htb]
	\begin{center}
		\includegraphics[width=8cm]{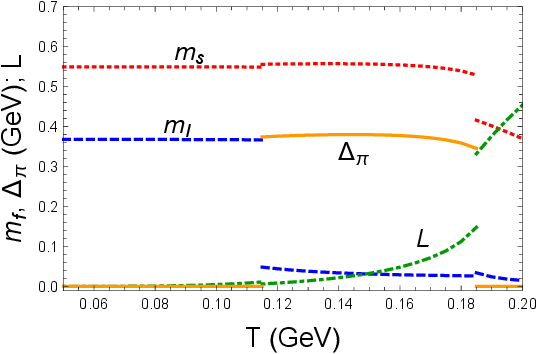}
		\caption{The quark masses $m_{\rm f}$, pion condensate $\Delta_\pi$, and Polyakov loop $L$ as functions of the temperature $T$, decreasing with the expansion of the early universe, for the case $eH=0.01\,{\rm GeV}^2$ and $l_{\rm e}^{\rm M}+l_\mu^{\rm M}=-0.2$. Here, $m_{\rm u}$ and $m_{\rm d}$ are almost the same with each other and thus can be consistently presented as $m_{\rm l}$.}\label{OrderPs}
	\end{center}
\end{figure}

\begin{figure}[!htb]
	\begin{center}
		\includegraphics[width=8cm]{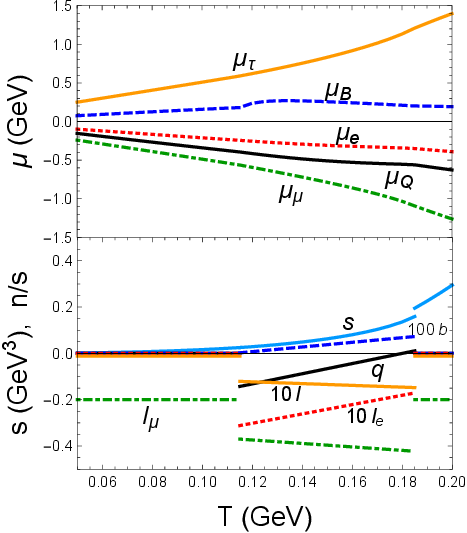}
		\caption{Upper panel: the cosmic trajectories of the chemical potentials of electric charge ($\mu_{\rm Q}$), baryon ($\mu_{\rm B}$), and lepton flavors ($\mu_{\rm e}, \mu_\mu$, and $\mu_\tau$) as functions of the temperature $T$; lower panel: the entropy density ($s$) and density-entropy ratios ($q, b, l_{\rm e},l_{\rm \mu}$, and $l$) as functions of the temperature $T$. The same as that in Fig.~\ref{OrderPs}, we consider the case $eH=0.01\,{\rm GeV}^2$ and $l_{\rm e}^{\rm M}+l_\mu^{\rm M}=-0.2$.}\label{mun}
	\end{center}
\end{figure}

Finally, we take the case $eH=0.01\,{\rm GeV}^2$ for example to show the evolution features of the order parameters -- quark masses $m_{\rm f}$, pion condensate $\Delta_\pi$, and Polyakov loop $L$ in Fig.\ref{OrderPs}. Since temperature decreases with time in the early universe, we call in the following the transition points with larger and smaller temperature as the first and second ones, respectively. Then, the region between the first and second transition points corresponds to the pion superfluidity phase as $\Delta_\pi\neq0$. As we can see, both chiral symmetry and $I_3$ isospin symmetry get broken through the formation of a large $\Delta_\pi$ at the first transition point. Then, the $I_3$ isospin symmetry becomes restored at the second transition point, where mainly $m_{\rm l}$ and $\Delta_\pi$ exchange their roles and thus does not cause many changes to thermodynamical quantities. Compared to that, the first transition is much stronger since $m_{\rm l}$ stays small but there are large gaps of $\Delta_\pi, m_{\rm s}$, and $L$. The corresponding cosmic trajectories of the chemical potentials and the entropy and number densities are demonstrated in the upper and lower panels of Fig.\ref{mun}, respectively. While the chemical potentials evolve continuously across the transition points as should be, there is a non-monotonic feature in the spectrum of $\mu_{\rm B}$, stronger than that in the vanishing $H$ limit~\cite{Cao:2021gfk}. According to the lower panel, $q$ and $l_{\rm e}$, which vanish in the chiral phases, are more sensitive to the second transition point, but all the left are more sensitive to the first one. This is another indication that the first phase transition is stronger.
\section{Part II: first-order phase transition with Polyakov-Quark-Meson model}\label{FOPT}
It is hard to study the first-order phase transition dynamics in the framework of the PNJL model that involves four- and six-quark interaction terms. Since it had been shown that the NJL model could be bosonized to a model with only meson degrees of freedom~\cite{Klevansky:1992qe}, we alternatively adopt the simple two-flavor PQM model~\cite{Schaefer:2006ds} to demonstrate the transition process in this section. Note that there are two reasons why we focus on the two-flavor rather than three-flavor PQM model: firstly, the effect of strange quarks is small for the exploration of pion superfluidity~\cite{Cao:2021gfk}; secondly, we could reduce the number of coupled equations of motions to facilitate numerical evaluations. In the model, the quark degrees of freedom can be integrated out to introduce the effects of temperature, chemical potentials, and magnetic field into the mesonic system, and further analysis of bubble dynamics could just follow those given in the pioneer works~\cite{Coleman:1977py,Callan:1977pt} for boson systems.

\subsection{Bubble dynamics }\label{BD}
In a background magnetic field, the Lagrangian density of the renormalizable two-flavor PQM model~\cite{Schaefer:2006ds} is given by
\begin{widetext}
\begin{eqnarray}
{\cal L}_{\rm PQM}&=&-{B^2\over 2}-V(L,L)+{1\over2}\Big[(\partial_\mu\sigma)^2+(\partial_\mu\pi^0)^2\Big]+{\cal D}^{\mu\dagger}{\pi}^+{\cal D}_\mu{\pi}^--{\lambda\over4}\left[\sigma^2+(\pi^0)^2+2{\pi}^+{\pi}^--\upsilon^2\right]^2+c~\sigma\nonumber\\
&&+\bar{\psi}\Big[i\slashed{D}\!-\! i\gamma^4\!\!\left(ig{\cal A}^4\!+\!Q_{\rm q}\mu_{\rm Q}\!+\!{\mu_{\rm B}\over3}\right)-g\Big(\sigma+i\gamma^5(\tau_3\pi^0+\tau_+\pi^++\tau_-\pi^-)\Big)\Big]\psi,
\end{eqnarray}
\end{widetext}
where $\psi(x)=(u(x),d(x))^T$ denotes the two-flavor quark field, the derivative for ${\pi}^-$ is ${\cal D}_\mu=\partial_\mu-i\,e A_\mu+ i\,\mu_{\rm Q}\delta_{\mu 0}$ and $\tau_\pm={1\over \sqrt{2}}(\tau_1\pm i\tau_2)$. The energy terms of the background magnetic field and Polyakov loop are the same as those in Eq.\eqref{LH}. The linear term $c~\sigma$ violates chiral symmetry explicitly and we can verify that the Lagrangian has exact $U_3(1)$ chiral symmetry in the chiral limit $c=0$. The model parameters of the mesonic sector $\lambda,\upsilon$, and $c$ are fixed by the sigma mass $m_\sigma=660~{\rm MeV}$, pion mass $m_\pi=138~{\rm MeV}$, and pion decay constant $f_\pi=93~{\rm MeV}$; and the quark-meson coupling constant is determined by $m_q^v\equiv gf_\pi=m_\sigma/2$ in the chiral symmetry breaking phase (for the stability of nucleons, $N_cm_q^v>m_N$)~\cite{Schaefer:2006ds}. 

By integrating over the quark degrees of freedom, the Lagrangian can be bosonized as
\begin{widetext}
\begin{eqnarray}
{\cal L}_{\rm PQM}&=&-{B^2\over 2}-V(L,L)+{1\over2}\Big[(\partial_\mu\sigma)^2+(\partial_\mu\pi^0)^2\Big]+{\cal D}^{\mu\dagger}{\pi}^+{\cal D}_\mu{\pi}^--{\lambda\over4}\left[\sigma^2+(\pi^0)^2+2{\pi}^+{\pi}^--\upsilon^2\right]^2+c\,\sigma\nonumber\\
&&+{\rm Tr}\ln\Big[i\slashed{D}_{\rm q}\!-\! i\gamma^4\!\!\left(ig{\cal A}^4\!+\!Q_{\rm q}\mu_{\rm Q}\!+\!{\mu_{\rm B}\over3}\right)-g\Big(\sigma+i\gamma^5(\tau_3\pi^0+\tau_+\pi^++\tau_-\pi^-)\Big)\Big].
\end{eqnarray}
In mean field approximation, the Gibbs free energy for the chiral phases and pion superfluidity can be easily evaluated as
\bea
\Omega_{\rm PQM}^{\rm M}&=&-{H^2\over 2}+V(L,L)+{\lambda\over4}(\sigma_{\rm l}^2-\upsilon^2)^2-c\,\sigma_{\rm l}-N_c\sum_{{\rm f}=u,d}\left\{-{1\over8\pi^2}\int_0^\infty {ds\over s^3}\left(e^{-m_{\rm f}^2s}-e^{-{m_{\rm f}^{\rm v}}^2s}\right)\left({q_{\rm f}eHs
	\over\tanh(q_{\rm f}eHs)}-1\right)-\right.\nonumber\\
&&\left.\!\!\!\!\!\!\!\!\!\!\!\!\!\!\!\!\!{1\over8\pi^2}\int_0^\infty {ds\over s^3}e^{-{m_{\rm f}^{\rm v}}^2s}\left({q_{\rm f}eHs
	\over\tanh(q_{\rm f}eHs)}-1-{1\over 3}(q_{\rm f}eHs)^2\right)+2T\sum_{\rm t=\pm}{|q_{\rm f}eH|\over 2\pi}\sum_{n=0}^\infty\alpha_{\rm n}\int_{-\infty}^\infty {\di  k_3\over2\pi}K_{\rm f}^{\rm t}(E_{\rm f}^{\rm n},L,T,\mu_{\rm Q},\mu_{\rm B})\right\},
\eea
and 
\bea
\!\!\!\!\!\!\Omega_{\rm PQM}^{\rm \pi SF}&=&V(L,L)+{\lambda\over4}\left[\sigma_{\rm l}^2+2{\Delta_\pi^2}-\upsilon^2\right]^2-\mu_{\rm Q}^2\Delta_\pi^2-c\,\sigma-2N_{\rm c}\int\!\!{\di^3k\over(2\pi)^3}\sum_{t=\pm}\left[\Big|\epsilon_{\rm l}(k)+t\,{\mu_{\rm Q}\over2}\Big|-\epsilon_{\rm l}(k)\right]\nonumber\\
&&-2T\!\!\int\!\!{\di^3k\over(2\pi)^3}\!\sum_{t,u=\pm}Fl\left(L,u,E_{\rm l}^{\rm t}(k),{\mu_{\rm Q}\!+\!2\mu_{\rm B}\over6}\right)
\eea
\end{widetext}
by following the schemes given in Sec.\ref{PNJLB} and Sec.\ref{PNJL}, respectively. Note that the quark vacuum terms with cutoff are dropped to avoid double counting and the last but one in $\Omega_{\rm \pi SF}$ guarantees its form for  $\Delta_\pi=0$ to be consistent with that of $\Omega_{\rm PQM}^{\rm M}$ in the vanishing $H$ limit. Here, we have taken $m_{\rm u}=m_{\rm d}=g\sigma_{\rm l}$ as the involved magnetic field is relatively small. As we can see, the effects of temperature, magnetic field, and chemical potentials can be conveniently introduced into the system through the quark loops at the mean field approximation level. 

\begin{figure}[!htb]
	\begin{center}
		\includegraphics[width=8cm]{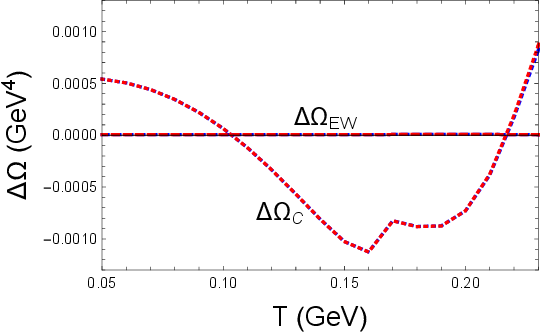}
		\caption{The free energy differences of strong interaction sector $\Delta\Omega_{\rm C}$ (dotted) and electroweak interaction sector $\Delta\Omega_{\rm EW}$ (dashed) as functions of temperature $T$ at $eH=0.01\,{\rm GeV^2}$. After taking the constraints $n^{\rm Q}=0, b^{\rm M}=8.6*10^{-11}, l^{\rm M}=-0.012, l^{\rm M}_{\rm e}=0,$ and $l^{\rm M}_{\rm \mu}=-0.2$ to fix the chemical potentials, we compare the cases with $H$-dependent (blue) and $H=0$ (red) fermion loops. At the low temperature $T=0.05\,{\rm GeV}$, the value $\Delta\Omega_{\rm C}=5.45\times10^{-4}\,{\rm GeV^4}$ actually equals the energy density of the background magnetic field $H^2/2$.}\label{dOmg}
	\end{center}
\end{figure}

The contributions of the electroweak interaction sector are the same as those given in Sec.\ref{EWB} and Sec.\ref{EW}. We compare the contributions of the strong and electroweak interaction sectors to the free energy difference between chiral phases and pion superfluidity in Fig.~\ref{dOmg} with the notations $\Delta\Omega_{\rm C}\equiv\Omega_{\rm PQM}^{\rm \pi SF}-\Omega_{\rm PQM}^{\rm M}$ and $\Delta\Omega_{\rm EW}\equiv\Omega_{\rm l}-\Omega_{\rm l}^{\rm M}$. According to our evaluations, $\Delta\Omega_{\rm EW}$ are two orders suppressed compared to $\Delta\Omega_{\rm C}$. We also present the results where the magnetic field $eH$ is artificially set to $0$ for all the charged fermion loops: since $|\Delta\Omega_{\rm C}-{H^2\over 2}-\Delta\Omega_{\rm C}|_{H\rightarrow0}|\ll {H^2\over 2}$, the magnetization from quark loops is negligible compared to the chosen magnetic field. 
Therefore, to conveniently study the first-order QCD phase transition, we can simply focus on the strong interaction sector and set $H=0$ in the quark loops for the chosen magnetic field. Actually, the latter approximation is consistent with that done in Eq.(48.18) for the study of superconductor in Ref.~\cite{Fetter2003b} and helps to explore bubble dynamics of the first phase transition. The bosonized Lagrangian density is now reduced to
\begin{eqnarray}
{\cal L}_{\rm PQM}&=&-{H^2\over 2}+{1\over2}(\partial_\mu\sigma)^2+{\cal D}^{\mu\dagger}{\pi}^+{\cal D}_\mu{\pi}^--U(L,\sigma,\pi^\pm),\nonumber
\end{eqnarray}
where the total potential is given by
\bea
U(L,\sigma,\pi^\pm)&\equiv& V(L,L)+{\lambda\over4}\left[\sigma^2+2{\pi}^+{\pi}^--\upsilon^2\right]^2-c\,\sigma\nonumber\\
&&-2N_{\rm c}\int\!\!{\di^3k\over(2\pi)^3}\sum_{t=\pm}\left[\Big|\epsilon(k)+t\,{\mu_{\rm Q}\over2}\Big|-\epsilon(k)\right]\nonumber\\
&&\!\!\!\!\!\!-2T\!\!\int\!\!{\di^3k\over(2\pi)^3}\!\sum_{t,u=\pm}\!\!Fl\left(L,u,E^{\rm t}(k),{\mu_{\rm Q}\!+\!2\mu_{\rm B}\over6}\right)\nonumber
\eea
with the energy functions $\epsilon(k)=\sqrt{k^2+g^2\sigma^2}$ and $E^{\pm}(k)=\sqrt{\left(\epsilon(k)\pm{\mu_{\rm Q}\over2}\right)^2+2g^2\pi^+\pi^-}$. Note that we have neglected the irrelevant $\pi^0$ degree of freedom and assumed local approximation~\cite{Cao:2018tzm} to reserve the space-time dependence of the fields $\sigma,\pi^\pm$, and $L$ in $U(L,\sigma,\pi^\pm)$.

By following the same scheme as in Sec.\ref{numerical}, the first phase transition point is found to be located at
\bea
&&\mu_{\rm Q}=-0.704\,{\rm GeV}, \mu_{\rm B}=0.352\,{\rm GeV}, \mu_{\rm e}=-0.437\,{\rm GeV}, \nonumber\\
&&\mu_{\rm \mu}=-1.372\,{\rm GeV}, \mu_{\rm \tau}=1.510\,{\rm GeV}, T=0.217\,{\rm GeV}
\eea
 with the homogeneous order parameters
 \bea
&&\sigma_{\rm l}=0.0026\,{\rm GeV}, L=0.509, \Delta_\pi=0\,{\rm GeV}; \nonumber\\
&&\sigma_{\rm l}=0.0018\,{\rm GeV}, L=0.154, \Delta_\pi=\pm0.116\,{\rm GeV}\label{solutions}
\eea
for the $\chi SR$ and pion superfluidity phases, respectively. Compared to the changing ratios of $\Delta_\pi$ and $L$, that of $\sigma_{\rm l}$ is small across the transition point; so we can safely drop $\sigma$ degree of freedom, as was done to $\pi^0$, for the bubble dynamics. Eventually, the Lagrangian density is further reduced to
\begin{eqnarray}
{\cal L}_{\rm PQM}=-{F_{\mu\nu}F^{\mu\nu}\over 4}\!+\!{\partial^\mu}\tilde{\pi}{\partial_\mu}\tilde{\pi}\!+\!(\mu_{\rm Q}^2\!-\!e^2{\bf A}^2)\tilde{\pi}^2\!-\!U(L,\tilde{\pi}),\nonumber\\\label{LPQM}
\end{eqnarray}
where we have set $\pi^-=\pi^+\equiv\tilde{\pi}$ without loss of generality and $U(L,\tilde{\pi})\equiv U(L,0,\tilde{\pi})$ should be understood. Note that a space-time dependent phase of $\pi^\pm$, that is, $\pi^\pm\equiv e^{\pm i\,\alpha(x)}\tilde{\pi}$, can be absorbed by redefinition of the vector potential $eA_\mu\rightarrow eA_{\mu}+\partial_\mu\alpha(x)$. And since ${\bf A}$ could be time-dependent in the bubble dynamics, we resort to a more general expression of the background EM field term by adopting $F_{\mu\nu}$. According to the opposite signs of $\mu_{\rm Q}^2$ and $e^2{\bf A}^2$ in the quadratic term of $\tilde{\pi}$, the effects of $\mu_{\rm Q}$ and $H$ could be intuitively expected to be opposite for the pion superfluidity -- they actually correspond to Bose-Einstein condensation and Meissner effects, respectively.

\begin{figure}[!htb]
	\begin{center}
		\includegraphics[width=8cm]{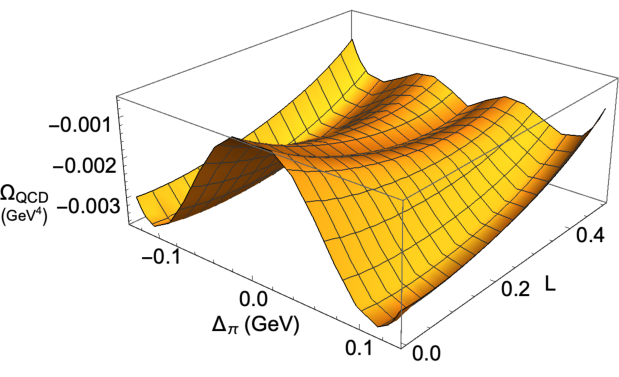}
		\caption{For homogeneous phases, the pure QCD part of the free energy $\Omega_{\rm QCD}$ is depicted as a function of the order parameters $\Delta_\pi$ and $L$ at temperature $T=0.2\,{\rm GeV}$.}\label{OMG_qcd}
	\end{center}
\end{figure}
In the mean field approximation with $\Delta_\pi\equiv\langle\tilde{\pi}\rangle$ and $L$ constants, the pure QCD part of the free energy, $\Omega_{\rm QCD}=-\mu_{\rm Q}^2\Delta_\pi^2+U(L,\Delta_\pi)$, is depicted as a function of $\Delta_\pi$ and $L$ in Fig.\ref{OMG_qcd} for $T=0.2\,{\rm GeV}$. Note that the temperature is a bit smaller than the first transition point $T_{\rm c}=0.217\,{\rm GeV}$ and then the bubble dynamics can be well explored in the transition from $\chi SR$ to pion superfluidity. Here, two kinds of minimal points can be identified, $(\Delta_\pi=0, L\sim0.4)$ and $(|\Delta_\pi|\sim0.1\,{\rm GeV}, L\sim0.1)$, which correspond to the $\chi SR$ and pion superfluidity phases, respectively. The sign of $\Delta_\pi$ is irrelevant, hence we take $ \Delta_\pi\geq 0$ in the following. By taking the magnetic part, $-{H^2/2}=-5.45\times10^{-4}\,{\rm GeV^4}$, into account in the $\chi SR$ phase, the gap between the free energies of these phases would reduce to $\varepsilon_{\rm v}=|\Delta\Omega|=7.28\times10^{-4}\,{\rm GeV^4}$.

\begin{figure}[!htb]
	\begin{center}
		\includegraphics[width=8cm]{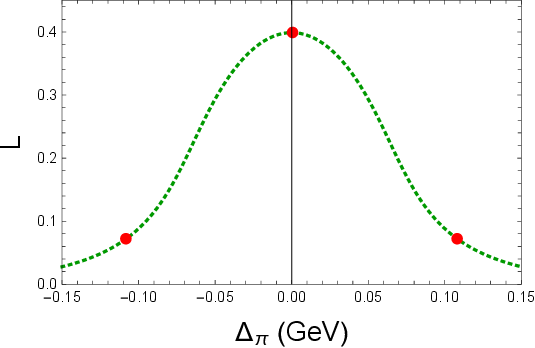}
		\caption{The solution ${L}$ as a function of $\tilde{\pi}$ from Eq.\eqref{EL} within the range where the homogeneous solutions Eq.\eqref{solutions} (red bullets) are covered.}\label{barL}
	\end{center}
\end{figure}

During the first-order phase transition, bubbles of true vacuum will be formed and then expand against the false vacuum~\cite{Coleman:1977py}. In such a case, the expectation values of the order parameters ($L$ and $\tilde{\pi}$) and vector potential ${\bf A}$ must be inhomogeneous across the space-time. By following the Euler-Lagrangian equation $\partial^\mu{\partial{\cal L}\over\partial (\partial^\mu \varphi)}-{\partial{\cal L}\over\partial \varphi}=0$, the corresponding equations of motion (EoMs) can be derived as
\begin{eqnarray}
\partial_{L}U(L,\tilde{\pi})&=&0,\label{EL}\\
2\partial^\mu\partial_\mu\tilde{\pi}+2(e^2{\bf A}^2-\mu_{\rm Q}^2)\tilde{\pi}+\partial_{\tilde{\pi}}U(L,\tilde{\pi})&=&0,\label{Epi}\\
\partial_0^2{\bf A}+\mathbb{\nabla}\times {\bf B}+2e^2\tilde{\pi}^2{\bf A}&=&0\label{eH}
\end{eqnarray}
with ${\bf B}\equiv\mathbb{\nabla}\times{\bf A}$. Note that we work in Coulomb gauge with $A_0=0$. As we can see, the effective pion field $\tilde{\pi}$ is the key to couple all the equations. Eq.\eqref{EL} is actually an algebra equation of $L$ and $\tilde{\pi}$ and can be solved numerically to obtain $L$ as a function of $\tilde{\pi}$, i.e. $L(\tilde{\pi})$, see the numerical results in Fig.\ref{barL}.
Substituting $L(\tilde{\pi})$ into Eq.\eqref{Epi}, we are then left with two coupled differential EoMs for $\tilde{\pi}$ and ${\bf A}$.
For homogeneous pion superfluidity with $\tilde{\pi}$ a nonzero constant, the static EoM of ${\bf B}$ can be obtained by taking curl of Eq.\eqref{eH} as
\bea
\mathbb{\nabla}^2 {\bf B}-2e^2\tilde{\pi}^2 {\bf B}&=&0.
\eea
The equation is similar to Eq.(49.11) in Ref.\cite{Fetter2003b}, so the Meissner effect resumes for $\tilde{\pi}\neq0$ in the relativistic case. Generally, the gauge coupling term $e^2{\bf A}^2\tilde{\pi}^2$ in Eq.\eqref{LPQM} favors $\tilde{\pi}\, {B}=0$ in the bulk for a homogeneous phase, so one can check that the solutions in Eq.\eqref{solutions} automatically apply to Eq.(\ref{EL}-\ref{eH}) as should be. Actually, with the temperature decreasing in the QCD epoch, the two solutions mainly correspond to the boundaries and centers of the bubbles in the first-order transition process. 

Now, the tough mission left is to work out the bubble structure from the coupled differential equations of motion Eqs.\eqref{Epi} and \eqref{eH}, that is, the forms of $\tilde{\pi}(t,{\bf r})$ and ${\bf A}(t,{\bf r})$, and thus $L(t,{\bf r})\equiv L(\tilde{\pi}(t,{\bf r}))$. To uniquely fix the bubble structure, we simply require the bubbles to smoothly approach to the $\chi SR$ phase at the boundary ${\bf r}\rightarrow\infty$. Firstly, as the background magnetic field is homogeneous and along $z$-direction, we assume the $B$ vortices as well as the $\tilde{\pi}$ bubbles to be cylindrical with their symmetry axes along $z$-direction. Hence, the bubbles are expected to be $z$ independent. Secondly, rotational symmetry would require $\tilde{\pi}({\bf r})=\tilde{\pi}(r)$ and $B_{\rm z}({\bf r})=B_{\rm z}(r)$, where $r$ is the radius in $x-y$ plane and the center is chosen to be at $x=y=0$. To make sure explicit rotational symmetry of Eq.\eqref{Epi}, we choose the symmetric gauge for the vector potential, that is, ${\bf A}(t,{\bf r})=-A(t,r)y\,{\bf i}+A(t,r)x\,{\bf j}$ and the corresponding magnetic field is $B_{\rm z}(t,r)={\partial\, r^2A(t,r)\over r\partial r}$. Within such a scheme, we are glad that the equations of motion of the $A_{\rm x}$ and $A_{\rm y}$ components reduce to the same one, and we have
\begin{eqnarray}
\!\!\!2\left[\partial_t^2\!-\!\partial_{\rm r}^2\!-\!{1\over r}\partial_{\rm r}\!+\!e^2A^2(r)r^2\!-\!\mu_{\rm Q}^2\right]\tilde{\pi}+\partial_{\tilde{\pi}}U(\tilde{\pi})&=&0,\label{Epitr}\\
\left(\partial_t^2-\partial_{\rm r}^2-{3\over r}\partial_{\rm r}+2e^2\tilde{\pi}^2\right)A(r)&=&0\label{eHtr}
\end{eqnarray}
instead of Eqs.\eqref{Epi} and \eqref{eH}. Note that we have used $U(\tilde{\pi})\equiv U(L(\tilde{\pi}),\tilde{\pi})$ for brevity here. 

To solve the coupled differential equations Eqs.\eqref{Epitr} and \eqref{eHtr}, boundary conditions are needed. Conventionally, to avoid singularities at the origins of space and time, we require the first order derivatives to be vanishing~\cite{Coleman:1977py}, that is, 
\bea
\!{\partial \tilde{\pi}\over\partial t}\Big|_{t=0}= {\partial \tilde{\pi}\over\partial r}\Big|_{r=0}={\partial A(r)\over\partial t}\Big|_{t=0}={\partial A(r)\over\partial r}\Big|_{r=0}=0.\label{BDs}
\eea
 As we can see, the magnetic effect is of order $o(r^2)$ around $r\sim0$ in Eq.\eqref{Epitr}, so we could neglect such term and solve the $O(3)$ symmetric equation of motion
\begin{eqnarray}
2\left[\partial_\rho^2+{2\over \rho}\partial_{\rho}+\!\mu_{\rm Q}^2\right]\tilde{\pi}-\partial_{\tilde{\pi}}U(\tilde{\pi})=0\label{Epi_r0}
\end{eqnarray}
 to get the imaginary time $\tau=i\,t$ dependence of $\tilde{\pi}$ at the origin $r=0$. Note that the Lorentz invariant solution was found to be with the lowest energy for a Lorentz invariant equation of motion. Then, it is natural to give the differential equation of motion for $\tilde{\pi}$ at the initial time $t=0$ as
 \begin{eqnarray}
2\left[\partial_r^2+{2\over r}\partial_{r}\!-\!e^2A^2(r)r^2+\!\mu_{\rm Q}^2\right]\tilde{\pi}-\partial_{\tilde{\pi}}U(\tilde{\pi})=0,
\label{Epi_t0}
\end{eqnarray}
which reduces to Eq.\eqref{Epi_r0} around $r\sim0$. Eq.\eqref{Epi_t0} is more useful than Eq.\eqref{Epi_r0} as the solution interpolates between the true and false vacua and could directly taken as the boundary condition for Eq.\eqref{Epitr}.

 Next, by taking curl of Eq.\eqref{eH}, we have
\begin{eqnarray}
&&\partial_0^2{\bf B}-\mathbb{\nabla}^2{\bf B}+2e^2\tilde{\pi}^2{\bf B}+4e^2\tilde{\pi}\mathbb{\nabla}\tilde{\pi}\times{\bf A}\nonumber\\
&=&[\partial_0^2{B}-\mathbb{\nabla}^2{B}+2e^2\tilde{\pi}^2{B}+4e^2\tilde{\pi}\mathbb\partial_{\rm r}\tilde{\pi}A(r)r]\hat{z}=0.\label{B_iso}
\end{eqnarray}
Due to the boundary conditions in Eq.\eqref{BDs}, the last term on the left-hand side of Eq.\eqref{B_iso} is of order $o(r^2)$ and can be safely dropped around $r\sim0$. Then, we get an $O(3)$ symmetric equation of motion for the magnetic field $B$, that is,
 \begin{eqnarray}
\partial_\rho^2B+{2\over\rho}\partial_\rho B-2e^2\tilde{\pi}^2B&=&0.\label{eH_r0}
\end{eqnarray}
Recalling the relation $B(t,r)={\partial\, r^2A(t,r)\over r\partial r}$, Eq.\eqref{BDs} implies similar boundary conditions for $B$: $${\partial B(t,r)\over\partial t}\Big|_{t=0}={\partial B(t,r)\over\partial r}\Big|_{r=0}=0.$$ So by substituting the solution $\tilde{\pi}(\rho)$ from Eq.\eqref{Epi_r0} into Eq.\eqref{eH_r0}, the form of $B(\rho)$ could be worked out and $A(\tau,r)$ follows as 
\bea
A(\tau,r)&=&{1\over r^2}\int_0^rB(\sqrt{\tau^2+s^2})sds\nonumber\\
&\approx&{1\over2} \left[B(\tau)+{1\over 2}{B'(\tau)\over \tau}{r}^2\right]
\eea
to order $o({r}^2)$. One can substitute $A(\tau,r)$ into Eq.\eqref{eH} to find it the same as Eq.\eqref{eH_r0} at $r=0$, so $A(\tau,0)={1\over2} B(\tau)$ at the spatial origin $r=0$. 
As mentioned, the form of $B(0,r)$ is more important. Following the previous discussions, it is natural to give the differential equation of motion for $B$ at the initial time from Eq.\eqref{B_iso} as
\begin{eqnarray}
\partial_r^2B_{\rm z}+{2\over r}\partial_r B_{\rm z}-2e^2\tilde{\pi}^2{B}-4e^2\tilde{\pi}\mathbb\partial_{\rm r}\tilde{\pi}A(r)r=0.\label{eH_t0}
\end{eqnarray}

\begin{figure}[!htb]
	\begin{center}
		\includegraphics[width=8cm]{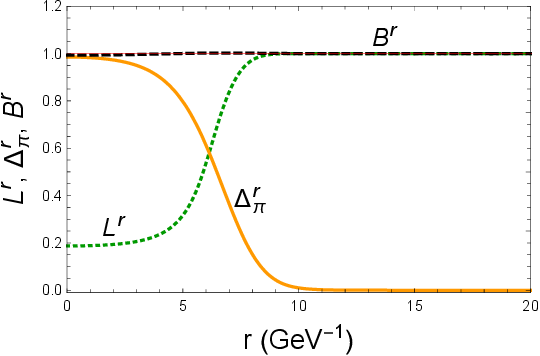}
		\caption{The solutions $\tilde{\pi}(r)$ (yellow solid), ${L}(r)$ (green dotted), and $B(r)$ (black dashed) as functions of the radius $r$ from Eqs.\eqref{EL}, \eqref{Epi_t0}, and \eqref{eH_t0}. The Polyakov loop and pion condensate are normalized by their maximal values to $L^{\rm r}$ and ${\Delta}_\pi^{\rm r}$, and the magnetic field $B$ by its background value $H$ to $B^{\rm r}$, respectively. The red baseline is $f(r)=1$.}\label{Field_r}
	\end{center}
\end{figure}

So by solving the differential equations $B(r)={\partial\, r^2A(r)\over r\partial r}$, Eq.\eqref{Epi_t0}, and Eq.\eqref{eH_t0} together, we could obtain the spatial structures of the fields $\tilde{\pi}, B$, and $A$ at the initial time $t=0$. The normalized results are presented in Fig.\ref{Field_r}. First of all, the bubble structures of $\tilde{\pi}$ and $L$ are almost the same as the case $B=0$ with the radius $R_b\approx 9.60\,{\rm GeV}^{-1}$, so it is a good approximation to treat the bubble formation with the Lorentz invariant assumption. Second, the characteristic length ($\xi\lesssim 6\,{\rm GeV}^{-1}$) of the pion condensate $\tilde{\pi}$ is consistent with that of the Polyakov loop $L$. Third, the magnetic field does not change much in the whole bubble range which means that the penetration depth is much larger than $\xi$, so the bubbles are quite like the vortices in Type-II superconductor~\cite{Fetter2003b}. Forth, a bump is found in the magnetic field around the bubble wall, which is more obvious when the temperature is closer to the transition point. Such a feature is a consequence of the latent heat released from the phase transition and the coupling between $\tilde{\pi}$ and $A$. The underlying physics is consistent with the logic how first-order phase transition could induce extremely large primordial magnetic field in the early universe~\cite{Vachaspati:1991nm}.

\begin{figure}[!htb]
	\begin{center}
		\includegraphics[width=8cm]{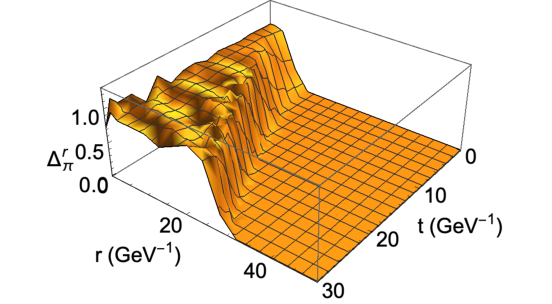}
		\includegraphics[width=8cm]{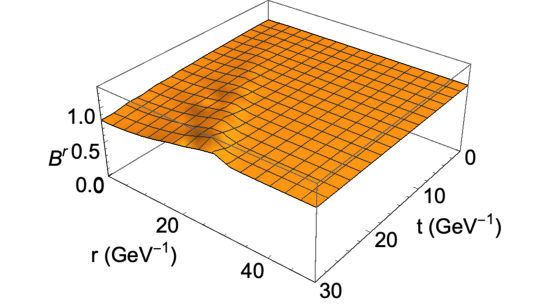}
		\caption{The space-time evolution of the reduced fields ${\tilde{\pi}}^{\rm r}$ (upper panel) and $B^{\rm r}$ (lower panel).}\label{Field_3D}
	\end{center}
\end{figure}

Now, we are ready to solve the full differential equations \eqref{Epi} and \eqref{eH} by adopting the initial conditions presented in Fig.\ref{Field_r}. The space-time evolution of the reduced fields $\tilde{\pi}^{\rm r}$ and $B^{\rm r}$ are illustrated in Fig.\ref{Field_3D}. The bubble wall of $\tilde{\pi}^{\rm r}$ expands $\sim 16\,{\rm GeV}^{-1} $ within the time interval $\sim 30\,{\rm GeV}^{-1} $, so the speed is smaller than the velocity of light as should be. Furthermore, with the development of the superconductor phase, the magnetic field decreases further deep inside the superconductor and would approach zero for a longer time. Meanwhile, the bump structure of the magnetic field becomes more and more sharp with time going, since more and more latent heat is released to the wall as the bubble expand. In this sense, the thin wall approximation could be adopted to calculate the nonvanishing quartic moment caused by bubble collisions, which is closely connected with the radiation of gravitational wave~\cite{Weinberg1972}.

\subsection{Gravitational wave within a toy system}\label{GWTS}
To explore bubble collisions, the tunneling probability density $\Gamma$ has to be estimated. Since our study has shown that the bubble formation is not far from 3-dimensional Lorentz invariant, $\Gamma$ could be evaluated in the $O(3)$ symmetric Euclidean space as $\Gamma={\cal A}\left({{\cal B}\over2\pi}\right)^{3/2}e^{-{\cal B}}$~\cite{Callan:1977pt}. In thin wall approximation, the involved coefficients are respectively
\begin{eqnarray}
{\cal A}&=&{(\mu R_b)^{7/3}\over R_b^3},\nonumber\\
{\cal B}&=&(2 R_b)4\pi \int_0^\infty\rho^2d\rho\left\{\left({d{\tilde\pi}\over d\rho}\right)^2+(e^2{A}^2(\rho)\rho^2-\mu_{\rm Q}^2){\tilde\pi}^2\right.\nonumber\\
&&\left.+U({\tilde\pi})-U(0)-{B^2(\rho)\!-\!H^2\over2}\right\},
\end{eqnarray}
 where $\mu$ is the renormalization scale for $2+1$ dimension~ \cite{Garriga:1993fh} and could be set to $\mu=1\,{\rm GeV}$ for QCD. Even though we have assumed the bubble dynamics to be $z$-independent to simplify the interplay between $B$ and $\tilde\pi$, the bubble must be bounded in $z$-direction in order to be consistent with the case $B$ is irrelevant. For the small bubbles, we set the size to be the same as that in $x$ and $y$ directions for simple estimation since Lorentz invariance is not violated in the $t-z$ space.  So the generated bubbles are roughly cylinders with the same height and diameter -- that is why $2 R_b$ shows up in the expression of ${\cal B}$. 
 
\begin{figure}[!htb]
	\begin{center}
		\includegraphics[width=8cm]{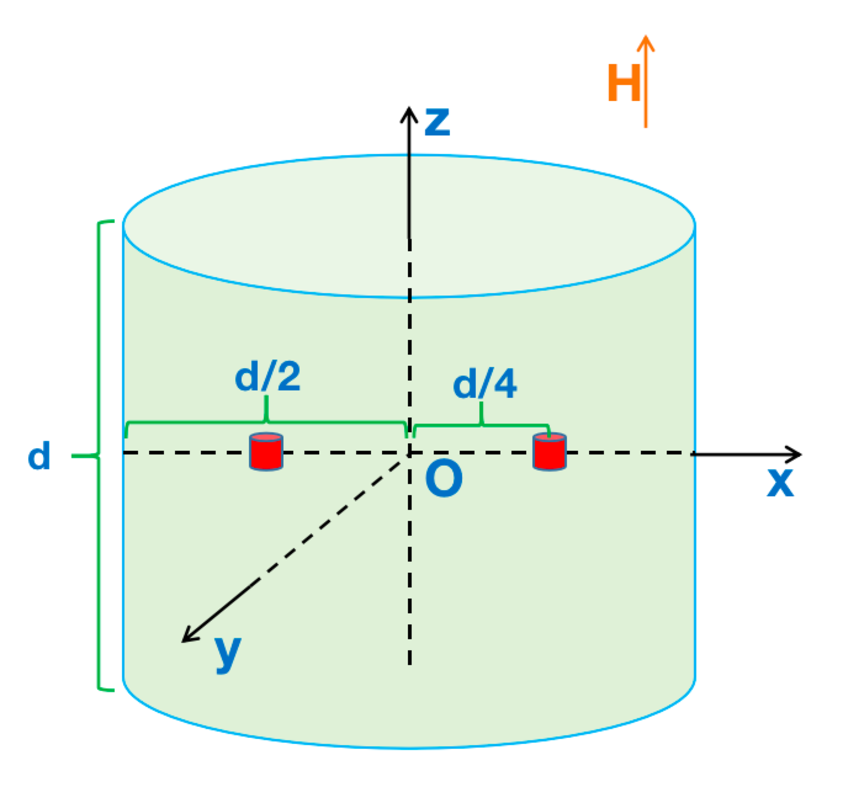}
		\caption{A toy model for bubble collision: the system and nucleated bubbles are all chosen to be cylinders with the height equal to the diameter and their axes along the direction of the background magnetic field, and the two bubbles are assumed to be generated simultaneously at the locations $x=\pm d/4,y=z=0$. The system size is $d=200\, {\rm GeV}^{-1}$ and the bubble radius was found to be $R_{\rm b}=9.60\, {\rm GeV}^{-1}$ according to the calculation in Sec.\ref{BD}.}\label{BC}
	\end{center}
\end{figure}

 Then, after inserting the solutions of Eqs.\eqref{EL}, \eqref{Epi_t0} and \eqref{eH_t0}, we find
 \begin{eqnarray}
{\cal A}=0.221\,{\rm GeV}^{-3},\ \
{\cal B}=14.2
\end{eqnarray}
and thus $\Gamma=5.29\times 10^{-7} {\rm GeV}^{-3}$. Then, considering a toy cylindrical system with height $h$ and diameter $d$ of the same size, that is, $h=d=200\, {\rm GeV}^{-1}$, we expect around two bubbles to be generated within the time scale $t\sim d/2$. In the following, we assume two bubbles to be generated simultaneously at $x=\pm d/4,y=z=0$ in such a system and study the features of the gravitational wave observed at the origin, see Fig.\ref{BC}. 
Compared to the common cases where the GW sources are far away from us~\cite{Weinberg1972}, we are assumed to be inside the source here as we are studying the magnetic effect in our own galaxy. Inspired by the fact that the characteristic frequency of GW does not depend on the bubble number in the early universe~\cite{Kosowsky:1992rz}, we expect the characteristic frequency to be determined by the energy scale at the QCD epoch and not sensitive to the details of bubble dynamics. In principle, the magnitude of the GW depends on the system size as the GW originated from different locations would overlap with each other. For larger system, we could divide it to several numbers of the toy system and the boundary of the toy system roughly corresponds to the overlapping region with the bubbles outside actually.

As the two bubbles expand and collide with each other, the variation of stress tensor would induce GW radiation, that is,
\begin{eqnarray}
T_{ij}({{\bf x}},t)=\partial_i\tilde{\pi}\partial_j\tilde{\pi}-{F_{i\nu} F^{j\nu}\over 2}.
\end{eqnarray}
The resultant GW strains are then given by~\cite{Weinberg1972}
\begin{eqnarray}\label{hij}
h_{+}(t)&=&{2G}\,\Re\int d\omega\int {dz\,d^2{\bf r}\over R}~ e^{-i\omega (t-R)}\!\left[T_{xx}\!-\!T_{yy}\right](z,{{\bf r}},\omega),\nonumber\\
h_{\times}(t)&=&{4G}\,\Re\int d\omega\int {dz\,d^2{\bf r}\over R}~ e^{-i\omega (t-R)}\!T_{xy}(z,{{\bf r}},\omega),\nonumber
\end{eqnarray}
where $\Re$ means the real part and $R$ is the distance of the GW source from the observer at the origin. For rough estimation, the strain will be evaluated by adopting the envelope approximation, which was shown to be in good agreement with the exact numerical evaluation~\cite{Kosowsky:1992vn}. The initial approximation is based on two simplifications: (i) the bubbles expand spherically with speed of light and do not interfere with each other; (ii) only the bubble walls that do not overlap with others and are inside the system are taken into account in the integration. Here, we modify (i) by requiring the bubbles to expand with half speed of light in $x-y$ plane according to the numerical results in Fig.\ref{Field_3D}. Hence, the GW strains become
\begin{eqnarray}\label{hij}
h_{+}(t)&=&{G}\,\Re\int d\omega\, e^{-i\omega t}H_{+}(\omega),\label{hp}\\
h_{\times}(t)&=&{G}\,\Re\int d\omega\, e^{-i\omega t}H_{\times}(\omega),
\end{eqnarray}
where $H_{+}(\omega)$ and $H_\times(\omega)$ depend on the stress tensors and are given in frequency space as
\bea
H_{+}(\omega)&=&{\varepsilon_{\rm v}\over2\pi}\!\int_0^\infty\!\!\!\!dt\,\!\int_{-l_{\rm z}(t)}^{l_{\rm z}(t)}\!\!\!dz\,r^2(t)\!\int_{S}\!d\theta{e^{i\omega [t+R(z,t,\theta)]}\over R(z,t,\theta)}  \cos(2\theta),\nonumber\\
\\
H_{\times}(\omega)&=&{\varepsilon_{\rm v}\over2\pi}\!\int_0^\infty\!\!\!\!dt\,\!\int_{-l_{\rm z}(t)}^{l_{\rm z}(t)}\!\!\!dz\,r^2(t)\!\int_{S}\!d\theta{e^{i\omega [t+R(z,t,\theta)]}\over R(z,t,\theta)} \sin(2\theta).\nonumber\\
\eea
Here, we define $l_{\rm z}(t)=R_{\rm b}+t$, $r(t)\approx R_{\rm b}+0.5\,t$, and $R(z,t,\theta)=\sqrt{z^2+r^2(t)+({d\over4})^2-\cos\theta\, r(t)\,{d\over2}}$, and $S$ means the surface of the non-overlapping bubble walls.

Since the configuration of bubbles is symmetric under the transformation $\theta\rightarrow-\theta$ at any time, we expect $H_{\times}(\omega)=0$. For $r(t)>{d\over2}$, the integration region of $\theta$ would start to be eaten by the overlapping with both the other bubble and the system boundary. Then, by taking into account the fact that the bubble configuration is reflectional symmetric with respect to all the axial planes, the explicit form of $H_{+}(\omega)$ is
\bea
H_{+}(\omega)&=&{2\varepsilon_{\rm v}\over\pi}\!\int\!dt\,r^2(t)\,\vartheta\left(r(t)-{d\over4}\right)\,\vartheta\left({d\over2}-l_{\rm z}(t)\right)\nonumber\\
&&\!\!\!\!\int_{-l_{\rm z}(t)}^{l_{\rm z}(t)}\!dz\int_{\theta_1}^{\theta_2} d\theta\,{e^{i\omega [t+R(z,t,\theta)]}\over R(z,t,\theta)} \cos(2\theta)\,\vartheta(\theta_2-\theta_1)\nonumber\\
&=&{2\varepsilon_{\rm v}\over\pi}\int\,dt\,r^2(t)\,{\rm Boole}\left({d\over4}<r(t)<\sqrt{5}{d\over4}\right)\nonumber\\
&&\!\!\!\!\!\!\!\!\vartheta\left({d\over2}-l_{\rm z}(t)\right)\!\int_{-l_{\rm z}(t)}^{l_{\rm z}(t)}\!dz\!\int_{\theta_1}^{\theta_2} \!\!d\theta{e^{i\omega [t+R(z,t,\theta)]}\over R(z,t,\theta)} \cos(2\theta).\nonumber\\\label{Hp1}
\eea
Here, the integral limits of $\theta$ are defined as $\cos\theta_1= {d\over 4r(t)}$ and $\cos\theta_2=-{3({d/4})^2-r^2(t)\over r(t){d/2}}$ according to the collisions with the other bubble and system boundary, respectively. Note that ${\rm Boole(conditions)}$ is the Boole function which is $1$ if the conditions are all true and $0$ otherwise.

Of course, after we obtained the explicit form of $H_{+}(\omega)$ by numerical calculations, $h_{+}$ can be evaluated according to the inverse Fourier transformation as shown in Eq.\eqref{hp}. On the other hand, we could also work out an integral form for $h_+$ by inserting Eq.\eqref{Hp1} into Eq.\eqref{hp} and then carry out numerical calculation directly. Actually, the integral over the frequency $\omega$ gives rise to a delta function $\delta(t-t'-R(z,t',\theta))$, which can be rewritten as ${R(z,t',\theta)\over r(t)\,{d/4}}\delta(\theta-\theta_3)\vartheta\left(t-t'\right)$ with $\cos\theta_3={z^2+r^2(t')+({d/4})^2-(t-t')^2\over r(t'){d/2}}$. Then, by completing the integration over $\theta$, we have
\begin{eqnarray}\label{hij}
h_{+}(t)&=&{4G\,\varepsilon_{\rm v}} \int\,dt'\,{r(t') \over {d/4}}{\rm Boole}\left({d\over4}<r(t')<\sqrt{5}{d\over4}\right)\nonumber\\
&&\vartheta\left({d\over2}-l_{\rm z}(t')\right)\vartheta\left(t-t'\right)\int_{-l_{\rm z}(t')}^{l_{\rm z}(t')}dz \,\cos (2\theta_3)\nonumber\\
&&{\rm Boole}\left(\theta_1<\theta_3<\theta_2\right),
\end{eqnarray}
 where one should keep in mind that $\theta_3$ depends on $t,t'$, and $z$. 
 
\begin{figure}[!htb]
	\begin{center}
		\includegraphics[width=8cm]{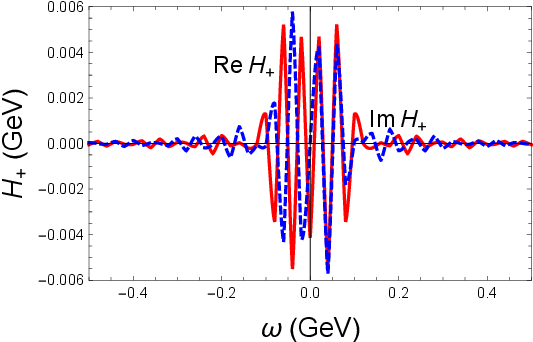}
		\includegraphics[width=8cm]{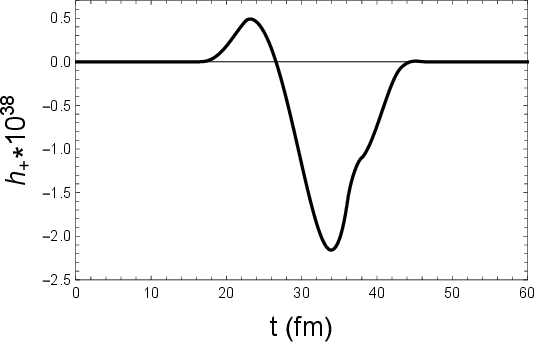}
		\caption{The spectra of the GW strain: the real (red solid) and imaginary (blue dashed) parts of $H_+$ with frequency $\omega$ in the upper panel and $h_+$ with time $t$ in the lower panel.}\label{GW}
	\end{center}
\end{figure}

 The numerical results for the GW strain are illustrated in Fig.\eqref{GW}. From the upper panel, we can identify that the real and imaginary parts of $H_+$ are even and odd functions of $\omega$, respectively, and the characteristic frequency is of order $0.1\,{\rm GeV}$. So the characteristic frequency is consistent with the energy scale of QCD, $\Lambda_{\rm QCD}=0.2$--$0.3\,{\rm GeV}$, and then that of the relic GW is of order $10^{-4}$--$10^{-5}\,{\rm eV}$ in our recent galaxy after been scaled by the factor $a=10^{12}$--$10^{12.5}$~\cite{Baumann:2022mni}. In other more familiar units, the relic frequency is of order $0.1$--$1\,{\rm K}$ or $10^9$--$10^{10}\,{\rm Hz}$, comparable to that of the cosmic microwave background. From the lower panel, we find the magnitude of the GW strain to be of order $10^{-38}$. Our system size is $5*10^{-13}\,{\rm fm}$ which corresponds to $0.5$--$5\, {\rm m}$ in recent universe. If the overlapping region of the GW is as large as $5\,{\rm km}$, the magnitude can be estimated to be of order $[(10^3)^3$--$10^4)^3]*10^{-38}=10^{-29}$--$10^{-26}$, quite within the capability of the next generation GW detectors~\cite{Evans:2016mbw}. If the bubble distribution is random, it is reasonable to assume that the characteristic frequency of $H_\times$ and magnitude of $h_\times$ should be of the same order as those of $H_+$ and $h_+$~\cite{Cao:2018tzm}.

\section{Summary}\label{summary}
In the first part, we have extended our study of the QCD phase diagram of the early universe~\cite{Cao:2021gfk} by taking into account primordial magnetic field within PNJL model. By referring to Gibbs free energy, we were able to demonstrate the Meissner effect at mean field level, that is, finite external magnetic field $H$ tends to reduce the free energy of chiral phases rather than that of superconducting pion superfluidity. Now, the transitions between chiral phases and pion superfluidity were well explored and were found to be of first order at relatively small $H$, compared to that of second order for vanishing $H$. With $H$ increasing, the regime of $\pi SF$ shrinks quickly and vanishes at the threshold value $eH=0.0222~{\rm GeV}^2$, so first-order transition to $\pi SF$ is only possible when the primordial magnetic field is around the lower limit of the estimated magnitude $10^{-2}$--$1\,{\rm GeV}^2$. Then, taking the case with $eH=0.01~{\rm GeV}^2$ and $l_{\rm e}^{\rm M}+l_\mu^{\rm M}=-0.2$ for example, we showed the the evolutions of order parameters, entropy, densities, and the cosmic trajectories of the chemical potentials with temperature, which decreases with time in the early universe. There, two first-order transitions were explicitly demonstrated according to the non-continuity of the order parameters, entropy, and densities; and the first transition was found to be stronger than the second one.

In the second part, we have adopted the two-flavor PQM model instead to study bubble dynamics during the first-order QCD transition. The equations of motion of mesons could be easily obtained in the PQM model and then we could conveniently apply the formalism of bubble dynamics~\cite{Coleman:1977py,Callan:1977pt} well developed before to the QCD system. At the first transition point, we reduced the set of EoMs by suppressing $\pi^0$ and $\sigma$ degrees of freedom, since the expectation value of $\pi^0$ is zero and that of $\sigma$ remains small across the point at mean field approximation. After working out the dependence of the Polyakov loop on charged pion condensate, we could eventually reduce the number of coupled EoMs to two, one for charged pion and the other for magnetic field. From that, we could directly reproduce the Meissner effect for the relativistic system and the expanding features of bubbles were well demonstrated. Then, the generating probability density was estimated for bubbles in the PQM model and we introduced a toy model by considering a cylindrical system where two bubbles could form. The two bubbles would expand and collide with each other to produce gravitational wave. By adopting the envelope approximation, the spectra of GW strain were briefly studied with respect to the variations of both frequency and time, and the characteristic frequency of the relic GW was estimated to be of the order $0.1$--$1\,{\rm K}$ or $10^9$--$10^{10}\,{\rm Hz}$ in our galaxy. For comparison, the characteristic frequency of the relic GW produced due to inflation~\cite{Starobinsky:1979ty} is of order $10^{-10}$--$10^{-8}\,{\rm Hz}$ when the early universe passed through the pion superfluidity phase with second-order transitions~\cite{Vovchenko:2020crk}

We have to admit that the recent study of gravitational wave generated from first-order QCD transition is quite preliminary, more realistic calculations should be carried out in the future. The dynamics with more bubbles would be considered for a much larger volume, and then the characteristic frequency and magnitude of GW strain can be further constrained. The contribution of domain walls to GW is also an interesting topic to explore just after the first-order QCD transition~\cite{Child:2012qg,Wei:2022poh}. On the other hand, the bubble structure shown in Fig.\ref{Field_r} seems to indicate that the Type-II superconducting pion superfluidity is more favored in an external magnetic field. It happens that the study of phonon modes favors such possibility~\cite{Adhikari:2022cks}, so we will check that by considering magnetic vortical structure in the future. Though the transitions involving Type-II superconductor are usually of second order, direct GW emission is still supposed to be generated from the collisions among vortices presented there.

\section*{Acknowledgement}
G.C. thanks Xian Gao for his help to give a reference for the scaling factor. G.C. is supported by the National Natural Science Foundation of China with Grant No. 11805290.


\begin{thebibliography}{99}

\bibitem{Weizsacker:1935bkz}
C.~F.~V.~Weizsacker,
``Zur Theorie der Kernmassen,''
Z. Phys. \textbf{96}, 431-458 (1935).

\bibitem{Hofstadter:1956qs}
R.~Hofstadter,
``Electron scattering and nuclear structure,''
Rev. Mod. Phys. \textbf{28}, 214-254 (1956).

\bibitem{Fetter2003a}
A. L. Fetter and J. D. Walecka, {\it Quantum theory of many-particle systems}, Chapter 11 (McGraw-Hill Book Company, New York, 1971).

\bibitem{Li:2008gp}
B.~A.~Li, L.~W.~Chen and C.~M.~Ko,
``Recent Progress and New Challenges in Isospin Physics with Heavy-Ion Reactions,''
Phys. Rept. \textbf{464}, 113-281 (2008).

\bibitem{Elgaroy:1996mg}
O.~Elgaroy, L.~Engvik, M.~Hjorth-Jensen and E.~Osnes,
``Superfluidity in beta stable neutron star matter,''
Phys. Rev. Lett. \textbf{77}, 1428-1431 (1996).

\bibitem{Ravenhall:1983uh}
D.~G.~Ravenhall, C.~J.~Pethick and J.~R.~Wilson,
``STRUCTURE OF MATTER BELOW NUCLEAR SATURATION DENSITY,''
Phys. Rev. Lett. \textbf{50}, 2066-2069 (1983).

\bibitem{Hashimoto:1984pap}
M.~a.~Hashimoto, H.~Seki and M.~Yamada,
``Shape of Nuclei in the Crust of Neutron Star,''
Prog. Theor. Phys. \textbf{71}, no.2, 320-326 (1984).

\bibitem{Glendenning:1992vb}
N.~K.~Glendenning,
``First order phase transitions with more than one conserved charge: Consequences for neutron stars,''
Phys. Rev. D \textbf{46}, 1274-1287 (1992).

\bibitem{Akmal:1997ft}
A.~Akmal and V.~R.~Pandharipande,
``Spin - isospin structure and pion condensation in nucleon matter,''
Phys. Rev. C \textbf{56}, 2261-2279 (1997).

\bibitem{Lee:1996ef}
C.~H.~Lee,
``Kaon condensation in dense stellar matter,''
Phys. Rept. \textbf{275}, 255-341 (1996).

\bibitem{Heiselberg:1999mq}
H.~Heiselberg and M.~Hjorth-Jensen,
``Phases of dense matter in neutron stars,''
Phys. Rept. \textbf{328}, 237-327 (2000).
 
\bibitem{Lee:1974ma}
T.~D.~Lee and G.~C.~Wick,
``Vacuum Stability and Vacuum Excitation in a Spin 0 Field Theory,''
Phys. Rev. D \textbf{9}, 2291-2316 (1974).

\bibitem{Lee:1974kn}
T.~D.~Lee,
``Abnormal Nuclear States and Vacuum Excitations,''
Rev. Mod. Phys. \textbf{47}, 267-275 (1975).

\bibitem{Yagi:2005yb}
K.~Yagi, T.~Hatsuda and Y.~Miake, {\it Quark-gluon plasma: From big bang to little bang}, (Cambridge University Press, Cambridge, 2005).

\bibitem{Aoki:2006we}
Y.~Aoki, G.~Endrodi, Z.~Fodor, S.~D.~Katz and K.~K.~Szabo,
``The Order of the quantum chromodynamics transition predicted by the standard model of particle physics,''
Nature \textbf{443}, 675-678 (2006).

\bibitem{Bhattacharya:2014ara}
T.~Bhattacharya, M.~I.~Buchoff, N.~H.~Christ, H.~T.~Ding, R.~Gupta, C.~Jung, F.~Karsch, Z.~Lin, R.~D.~Mawhinney and G.~McGlynn, \textit{et al.}
``QCD Phase Transition with Chiral Quarks and Physical Quark Masses,''
Phys. Rev. Lett. \textbf{113}, no.8, 082001 (2014).

\bibitem{Floris:2014pta}
M.~Floris,
``Hadron yields and the phase diagram of strongly interacting matter,''
Nucl. Phys. A \textbf{931}, 103-112 (2014).

\bibitem{Adamczyk:2017iwn}
L.~Adamczyk \textit{et al.} [STAR],
``Bulk Properties of the Medium Produced in Relativistic Heavy-Ion Collisions from the Beam Energy Scan Program,''
Phys. Rev. C \textbf{96}, no.4, 044904 (2017).

\bibitem{Luo:2017faz}
X.~Luo and N.~Xu,
``Search for the QCD Critical Point with Fluctuations of Conserved Quantities in Relativistic Heavy-Ion Collisions at RHIC : An Overview,''
Nucl. Sci. Tech. \textbf{28}, no.8, 112 (2017).

\bibitem{Baym:2017whm}
G.~Baym, T.~Hatsuda, T.~Kojo, P.~D.~Powell, Y.~Song and T.~Takatsuka,
``From hadrons to quarks in neutron stars: a review,''
Rept. Prog. Phys. \textbf{81}, no.5, 056902 (2018).

\bibitem{Witten:1984rs}
E.~Witten,
``Cosmic Separation of Phases,''
Phys. Rev. D \textbf{30}, 272-285 (1984).

\bibitem{Alford:2007xm}
M.~G.~Alford, A.~Schmitt, K.~Rajagopal and T.~Sch\"afer,
``Color superconductivity in dense quark matter,''
Rev. Mod. Phys. \textbf{80}, 1455-1515 (2008).

\bibitem{Fukushima:2015bda}
K.~Fukushima and T.~Kojo,
``The Quarkyonic Star,''
Astrophys. J. \textbf{817}, no.2, 180 (2016).

\bibitem{McLerran:2018hbz}
L.~McLerran and S.~Reddy,
``Quarkyonic Matter and Neutron Stars,''
Phys. Rev. Lett. \textbf{122}, no.12, 122701 (2019).

\bibitem{Cao:2020byn}
G.~Cao and J.~Liao,
``A field theoretical model for quarkyonic matter,''
JHEP \textbf{10}, 168 (2020).

\bibitem{Cao:2022inx}
G.~Cao,
``Quarkyonic matter state of neutron stars,''
Phys. Rev. D \textbf{105}, no.11, 114020 (2022).

\bibitem{Baumann:2022mni}
D.~Baumann, {\it Cosmology}, Chapter 3 (Cambridge University Press, Cambridge, 2022).

\bibitem{Trodden:1998ym}
M.~Trodden,
``Electroweak baryogenesis,''
Rev. Mod. Phys. \textbf{71}, 1463-1500 (1999).

\bibitem{Vachaspati:1991nm}
T.~Vachaspati,
``Magnetic fields from cosmological phase transitions,''
Phys. Lett. B \textbf{265}, 258-261 (1991).

\bibitem{Son:1998my}
D.~T.~Son,
``Magnetohydrodynamics of the early universe and the evolution of primordial magnetic fields,''
Phys. Rev. D \textbf{59}, 063008 (1999).

\bibitem{Grasso:2000wj}
D.~Grasso and H.~R.~Rubinstein,
``Magnetic fields in the early universe,''
Phys. Rept. \textbf{348}, 163-266 (2001).

\bibitem{Planck:2015fie}
P.~A.~R.~Ade \textit{et al.} [Planck],
``Planck 2015 results. XIII. Cosmological parameters,''
Astron. Astrophys. \textbf{594}, A13 (2016).


\bibitem{Vovchenko:2020crk}
V.~Vovchenko, B.~B.~Brandt, F.~Cuteri, G.~Endr\H{o}di, F.~Hajkarim and J.~Schaffner-Bielich,
``Pion Condensation in the Early universe at Nonvanishing Lepton Flavor Asymmetry and Its Gravitational Wave Signatures,''
Phys. Rev. Lett. \textbf{126}, no.1, 012701 (2021).

\bibitem{Middeldorf-Wygas:2020glx}
M.~M.~Middeldorf-Wygas, I.~M.~Oldengott, D.~B\"odeker and D.~J.~Schwarz,
``Cosmic QCD transition for large lepton flavor asymmetries,''
Phys. Rev. D \textbf{105}, no.12, 123533 (2022).

\bibitem{Cao:2021gfk}
G.~Cao, L.~He and P.~Zhang,
``Reentrant pion superfluidity and cosmic trajectories within a PNJL model,''
Phys. Rev. D \textbf{104}, no.5, 054007 (2021).

\bibitem{Fetter2003b}
A. L. Fetter and J. D. Walecka, {\it Quantum theory of many-particle systems}, Chapter 13 (McGraw-Hill Book Company, New York, 1971).

\bibitem{Hogan:1986qda}
C.~J.~Hogan,
``Gravitational radiation from cosmological phase transitions,''
Mon. Not. Roy. Astron. Soc. \textbf{218}, 629-636 (1986).

\bibitem{Kosowsky:1992rz} 
  A.~Kosowsky, M.~S.~Turner and R.~Watkins,
  ``Gravitational waves from first order cosmological phase transitions,''
  Phys.\ Rev.\ Lett.\  {\bf 69}, 2026 (1992).
  
\bibitem{Child:2012qg}
H.~L.~Child and J.~T.~Giblin, Jr.,
``Gravitational Radiation from First-Order Phase Transitions,''
JCAP \textbf{10}, 001 (2012).
  
\bibitem{Lewicki:2020azd}
M.~Lewicki and V.~Vaskonen,
``Gravitational waves from colliding vacuum bubbles in gauge theories,''
Eur. Phys. J. C \textbf{81}, no.5, 437 (2021)
[erratum: Eur. Phys. J. C \textbf{81}, no.12, 1077 (2021)].
  
\bibitem{Wei:2022poh}
D.~Wei and Y.~Jiang,
``Domain wall networks from first-order phase transitions and gravitational waves,''
[arXiv:2208.07186 [hep-ph]].

\bibitem{Pippard1953}
A. B. Pippard, "An Experimental and Theoretical Study of the Relation between Magnetic Field and Current in a Superconductor," Proc. Roy. Soc. (London), A216:547 (1953).

\bibitem{London1935}
F. London and H. London, "The electromagnetic equations of the supraconductor," Proc. Roy.  Soc. (London), A149:71 (1935).


\bibitem{Fukushima:2017csk}
K.~Fukushima and V.~Skokov,
``Polyakov loop modeling for hot QCD,''
Prog. Part. Nucl. Phys. \textbf{96}, 154-199 (2017).

\bibitem{Klevansky:1992qe}
  S.~P.~Klevansky,
  ``The Nambu-Jona-Lasinio model of quantum chromodynamics,''
  Rev.\ Mod.\ Phys.\  {\bf 64}, 649 (1992).

\bibitem{Hatsuda:1994pi} 
  T.~Hatsuda and T.~Kunihiro,
  ``QCD phenomenology based on a chiral effective Lagrangian,''
  Phys.\ Rept.\  {\bf 247}, 221 (1994).
  
\bibitem{Cao:2021rwx}
G.~Cao,
``Recent progresses on QCD phases in a strong magnetic field: views from Nambu\textendash{}Jona-Lasinio model,''
Eur. Phys. J. A \textbf{57}, no.9, 264 (2021).
  
\bibitem{Cao:2023bmk}
G.~Cao and J.~Li,
``A self-consistent thermodynamic potential for a magnetized QCD matter,''
[arXiv:2301.04308 [hep-ph]].

 
 \bibitem{Kapusta2006} J. I. Kapusta and C. Gale, {\it Finite-Temperature Field Theory: Principles and Applications} (Cambridge University Press, Cambridge, 2006).
 
\bibitem{Schwinger:1951nm}
J.~S.~Schwinger,
``On gauge invariance and vacuum polarization,''
Phys. Rev. \textbf{82}, 664-679 (1951).

\bibitem{Son:2000xc}
D.~T.~Son and M.~A.~Stephanov,
``QCD at finite isospin density,''
Phys. Rev. Lett. \textbf{86}, 592-595 (2001).
   
\bibitem{Zhuang:1994dw}
P.~Zhuang, J.~Hufner and S.~P.~Klevansky,
``Thermodynamics of a quark - meson plasma in the Nambu-Jona-Lasinio model,''
Nucl.\ Phys.\ A {\bf 576}, 525 (1994).

\bibitem{Rehberg:1995kh}
P.~Rehberg, S.~P.~Klevansky and J.~Hufner,
``Hadronization in the SU(3) Nambu-Jona-Lasinio model,''
Phys.\ Rev.\ C {\bf 53}, 410 (1996).

\bibitem{Oldengott:2017tzj}
I.~M.~Oldengott and D.~J.~Schwarz,
``Improved constraints on lepton asymmetry from the cosmic microwave background,''
EPL \textbf{119}, no.2, 29001 (2017).

\bibitem{Verschuur1974} G.L. Verschuur and K.I. Kellermann, {\it Galactic and Extragalactic Radio Astronomy} (Springer-Verlag, Berlin, 1974).

\bibitem{Schaefer:2006ds}
B.~J.~Schaefer and J.~Wambach,
``Susceptibilities near the QCD (tri)critical point,''
Phys. Rev. D \textbf{75} (2007), 085015.

\bibitem{Coleman:1977py} 
S.~R.~Coleman,
``The Fate of the False Vacuum. 1. Semiclassical Theory,''
Phys.\ Rev.\ D {\bf 15}, 2929 (1977)
Erratum: [Phys.\ Rev.\ D {\bf 16}, 1248 (1977)].
	
\bibitem{Callan:1977pt} 
C.~G.~Callan, Jr. and S.~R.~Coleman,
``The Fate of the False Vacuum. 2. First Quantum Corrections,''
Phys.\ Rev.\ D {\bf 16}, 1762 (1977).

\bibitem{Cao:2018tzm}
G.~Cao and S.~Lin,
``Gravitational Wave from Phase Transition inside Neutron Stars,''
[arXiv:1810.00528 [nucl-th]].

 \bibitem{Weinberg1972}S. Weinberg, {\it Gravitation and Cosmology} (Wiley, New York, 1972). 

\bibitem{Garriga:1993fh} 
  J.~Garriga,
  ``Nucleation rates in flat and curved space,''
  Phys.\ Rev.\ D {\bf 49}, 6327 (1994).


\bibitem{Kosowsky:1992vn} 
  A.~Kosowsky and M.~S.~Turner,
  ``Gravitational radiation from colliding vacuum bubbles: envelope approximation to many bubble collisions,''
  Phys.\ Rev.\ D {\bf 47}, 4372 (1993).
  
 
\bibitem{Evans:2016mbw} 
B.~P.~Abbott {\it et al.} [LIGO Scientific Collaboration],
``Exploring the Sensitivity of Next Generation Gravitational Wave Detectors,''
Class.\ Quant.\ Grav.\  {\bf 34}, no. 4, 044001 (2017).

\bibitem{Starobinsky:1979ty}
A.~A.~Starobinsky,
``Spectrum of relict gravitational radiation and the early state of the universe,''
JETP Lett. \textbf{30}, 682-685 (1979).

\bibitem{Adhikari:2022cks}
P.~Adhikari, E.~Leeser and J.~Markowski,
``Phonon modes of magnetic vortex lattices in finite isospin QCD,''
[arXiv:2205.13369 [hep-ph]].

\end{thebibliography}
\end{document}